\documentclass[
  aps,
  prl,
  reprint,
  10pt,
  superscriptaddress,
  footinbib,
  amsfonts,
  amsmath,
  amssymb
]{revtex4-2}

\usepackage{bm}

\usepackage{microtype}
\usepackage{graphicx}
\usepackage{hyperref}
\usepackage{xcolor}
\hypersetup{
  pdftitle={Spin-selective tunneling in altermagnetic quantum dots},
  pdfauthor={Andrea Maiani, Simon Wozny, Jos\'e Carlos Abadillo-Uriel, Alberto Cortijo, Ram\'on
    Aguado, Rub\'en Seoane Souto},
  colorlinks,
  linkcolor={blue!80!black},
  citecolor={blue!80!black},
  urlcolor={blue!80!black}
}

\DeclareMathOperator{\tr}{tr}
\DeclareMathOperator{\dd}{d}

\begin{document}

\title{Spin-selective tunneling in altermagnetic quantum dots}

\author{Andrea Maiani}
\affiliation{Nordita, KTH Royal Institute of Technology, and Stockholm University, Hannes Alfv\'ens v\"ag 12, SE-10691 Stockholm, Sweden}

\author{Simon Wozny}
\affiliation{Department of Engineering, Reykjavik University, Menntavegur 1, IS-102 Reykjav{\'i}k, Iceland}

\author{Jos\'e Carlos Abadillo-Uriel}
\affiliation{Quantum Advanced Research Center (QuARC), Consejo Superior de Investigaciones Cient{\'i}ficas (CSIC), Sor Juana In{\'e}s de la Cruz 3, 28049 Madrid, Spain}
\affiliation{Instituto de Ciencia de Materiales de Madrid (ICMM), Consejo Superior de Investigaciones Cient{\'i}ficas (CSIC), Sor Juana In{\'e}s de la Cruz 3, 28049 Madrid, Spain}

\author{Alberto Cortijo}
\affiliation{Instituto de Ciencia de Materiales de Madrid (ICMM), Consejo Superior de Investigaciones Cient{\'i}ficas (CSIC), Sor Juana In{\'e}s de la Cruz 3, 28049 Madrid, Spain}

\author{Ram{\'o}n Aguado}
\affiliation{Quantum Advanced Research Center (QuARC), Consejo Superior de Investigaciones Cient{\'i}ficas (CSIC), Sor Juana In{\'e}s de la Cruz 3, 28049 Madrid, Spain}
\affiliation{Instituto de Ciencia de Materiales de Madrid (ICMM), Consejo Superior de Investigaciones Cient{\'i}ficas (CSIC), Sor Juana In{\'e}s de la Cruz 3, 28049 Madrid, Spain}

\author{Rub\'en Seoane Souto}
\affiliation{Quantum Advanced Research Center (QuARC), Consejo Superior de Investigaciones Cient{\'i}ficas (CSIC), Sor Juana In{\'e}s de la Cruz 3, 28049 Madrid, Spain}
\affiliation{Instituto de Ciencia de Materiales de Madrid (ICMM), Consejo Superior de Investigaciones Cient{\'i}ficas (CSIC), Sor Juana In{\'e}s de la Cruz 3, 28049 Madrid, Spain}

\date{November 2026}

\begin{abstract}
  Spin-based electronic devices rely on the generation and manipulation of spin-polarized
  currents, where spin selectivity is a key ingredient. In this work, we demonstrate
  that a quantum dot (QD) defined in an altermagnet acts as an ideal spin-selective element: the
  two spins are confined differently, which results in unequal tunnel amplitudes to the leads.
  Unlike ferromagnets, QDs in altermagnets allow for an electrical tuning of the spin splitting and
  spin selectivity through gates controlling the confining potential. The spin-dependent
  transmissions produce current steps of unequal height, providing a direct measure
  of spin selectivity. These effects persist in the presence of Coulomb interactions
  and spin--orbit coupling, which additionally enables coherent pseudospin rotations.
  These results establish altermagnetic QDs as versatile building blocks for
  gate-controlled spintronics and as sensitive probes for altermagnetism at the nanoscale.
\end{abstract}

\maketitle

\paragraph{Introduction.}

Spin-polarized currents are at the heart of spintronics, which aims at employing the electron spin
as a resource for fast and efficient electronics. Ferromagnetic materials play a central role in the
field, as they polarize currents by favoring or blocking the transmission of one of the
electron spins~\cite{Meservey_1994_Spin, Zutic_2004_Spintronics, Fert_2008_Nobel}. However,
their spin-filtering properties are determined by the intrinsic electronic band
structure and interfacial characteristics, which constrain the achievable spin
polarization and the external tunability.

\begin{figure}[h]
  \centering
  \includegraphics[width=0.85\columnwidth]{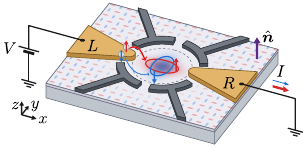}
  \caption{
    Sketch of the altermagnetic quantum dot (QD). Confinement gates (gray) define the dot in
    a two-dimensional altermagnet. Two metallic leads $L$ and $R$ contact the QD along $x$ and carry
    the current $I$ driven by the bias $V$. Red and blue contours: ground-state densities of the two
    N\'eel-spin species, elongated along $x$ and $y$ respectively, so that a contact on the $x$ axis
    couples more strongly to the up spin. Up (red) and down (blue) electrons tunnel from $L$ into
    their orbitals, the down spin at a suppressed rate, so the current is carried mostly by the up
    spin (thick red and thin blue arrows).
  }
  \label{fig:1}
\end{figure}

Altermagnets offer a new route for spin-based devices. In these collinear compensated magnets,
opposite-spin sublattices are related by crystal rotations, giving rise to a momentum-dependent
spin splitting despite their vanishing net magnetizations~\cite{Smejkal_2022_Emerging,
Mazin_2022_Editorial, Smejkal_2022_Giant, Hayami_2019_Momentum, Ahn_2019_Antiferromagnetism,
Naka_2019_Spin}. Their characteristic electronic structure leads to field-free spin currents,
magnetoresistance, Hall response, and spin-transistor
action~\cite{Gonzalezhernandez_2021_Efficient, Naka_2019_Spin, Naka_2020_Anomalous,
Smejkal_2022_Giant, Liu_2025_Altermagnetic, Zarzuela_2025_Transport, Sigales_PRB2026,
Mencos_Nature2026}, with a growing set of candidate materials, many of them
two-dimensional~\cite{Bai_2024_Altermagnetism, Song_2025_Altermagnets}. However, the absence of a
net magnetic moment complicates conventional magnetometric
detection~\cite{Baltz_2018_Antiferromagnetic, Amin_2024_Nanoscale}, while the
momentum-dependent spin splitting calls for probes sensitive to its anisotropy. Developing local,
all-electrical approaches to access and exploit altermagnets' spin structure remains
an important challenge.

Quantum dots (QDs) offer such an all-electrical control. These nanostructures have
discrete electronic states that can be controlled at the single charge
level~\cite{Hanson_2007_Spins, Reimann_2002_Electronic}. This degree of control has made it
possible, for example, to generate single-electron current
pulses~\cite{Kouwenhoven_1991_Quantized, Blumenthal_2007_Gigahertz, Pekola_2013_Single} or
control the quantum mechanical properties of individual
electrons~\cite{Loss_1998_Quantum, Elzerman_2004_Single, Burkard_2023_Semiconductor}. In
nonmagnetic hosts, time-reversal symmetry makes QD properties spin-independent and
spin selectivity requires magnetic fields or ferromagnetic contacts~\cite{Recher_2000_Quantum,
Hanson_2004_Semiconductor}, resulting in limited external control.

In this Letter, we demonstrate that QDs in semiconducting altermagnets are ideal
candidates for spin filtering. Their low carrier density allows us to define and control the
carrier confinement at the nanoscale, providing a means for rapidly tunable spin-polarized currents.
Quantum confinement transforms the momentum-dependent spin splitting into anisotropic
spatial profiles for opposite-spin wavefunctions, giving rise to spin-dependent
tunneling rates and, consequently, spin-selective transport. The shape and depth of
the confining potential determine the level-dependent spin splitting and spin-selective tunneling,
enabling electrical control over the spin polarization of the flowing current. These
conclusions remain valid in the presence of Coulomb repulsion and spin--orbit
coupling, which allows spin precession to be probed without an applied field.

Therefore, our work establishes a direct connection between altermagnets' electronic
structure and electrically controllable spin properties in confined geometries,
enabling the detection of altermagnetism at the nanoscale and a novel generation of
all-electrically controlled spin qubits~\cite{Abadillouriel_2026_All, Vakili_2026_Gate}.

\begin{figure}[h]
  \centering
  \includegraphics[width=1.0\columnwidth]{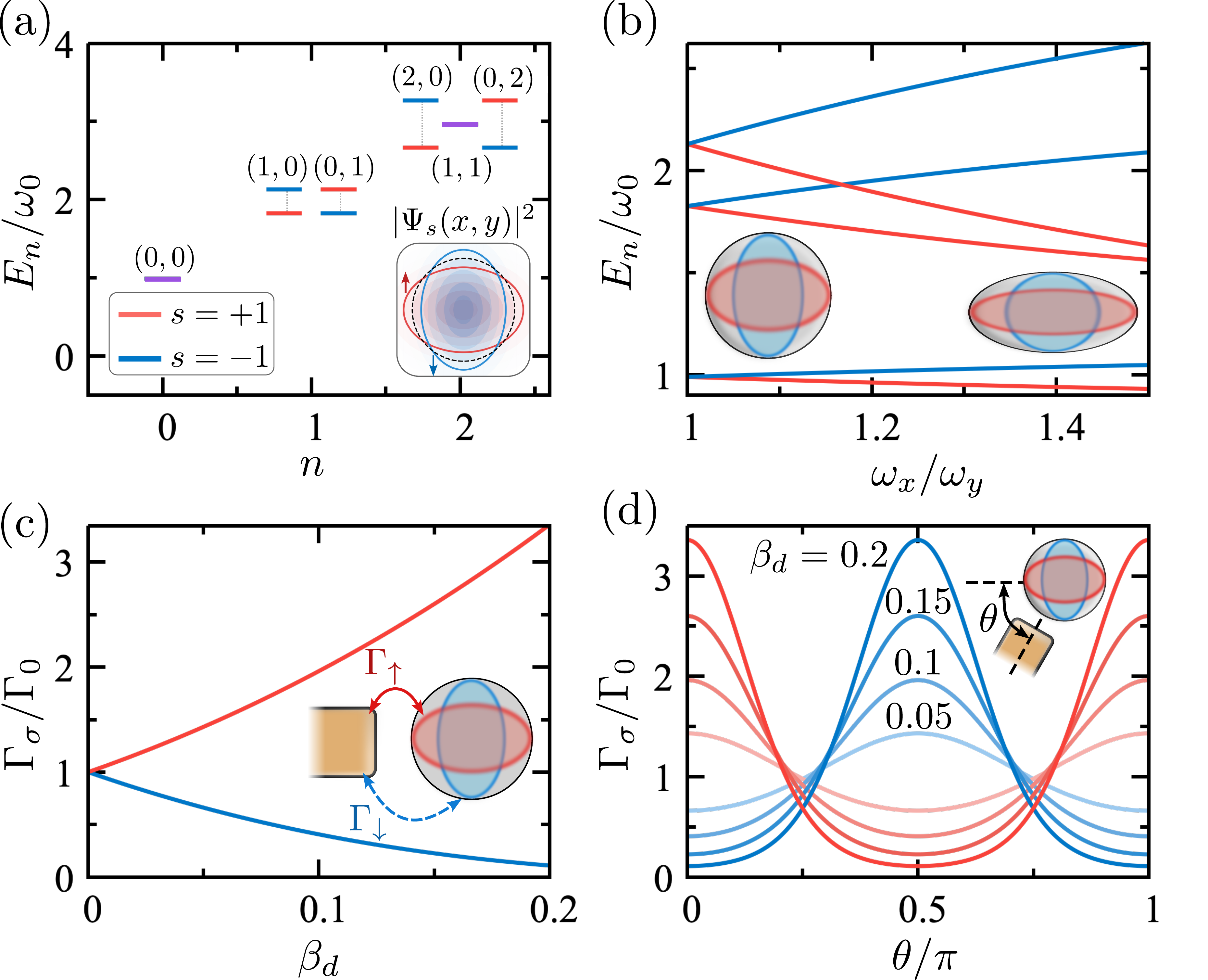}
  \caption{
    (a) Level energies of a circular altermagnetic QD for $\beta_d = 0.15$. The inset illustrates
    the different spatial extension of the lowest orbitals. (b) Spectrum of the altermagnetic
    QD as a function of the QD ellipticity. (c) Tunneling rate for up/down spin (red/blue line) as a
    function of the altermagnetic order $\beta_d$ for $\theta = 0$. (d) Tunneling rate as a
    function of the angle with respect to the altermagnetic axis, for different $\beta_d$ values.
    Parameters: $\omega_0 = 2$, $m = 1$, $R_{\mathrm{lead}} = 2$, and $\varphi = 0.25$.
  }
  \label{fig:2}
\end{figure}

\begin{figure}[h]
  \centering
  \includegraphics[width=1.0\columnwidth]{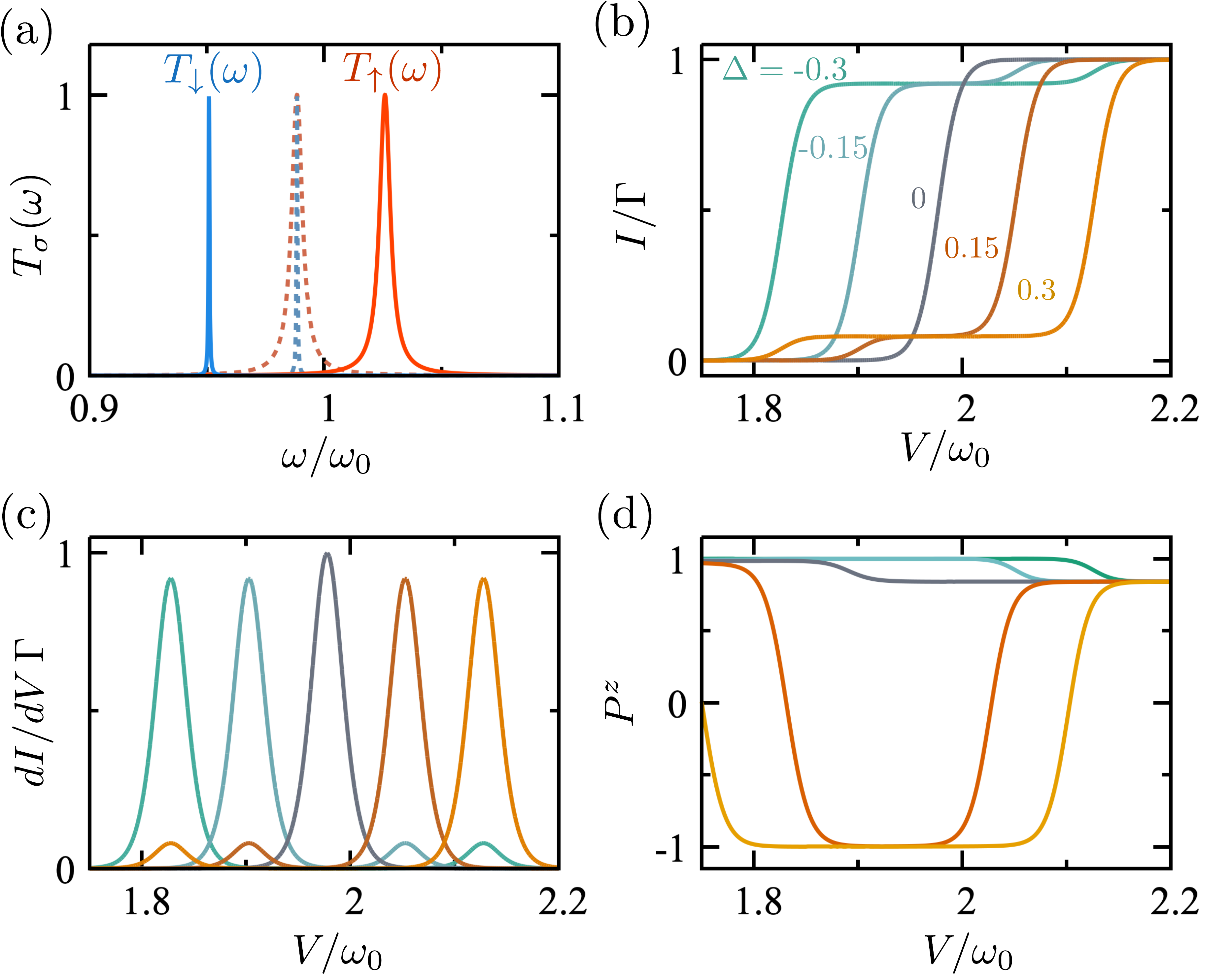}
  \caption{
    Spin-resolved transport through the altermagnetic dot. (a) Transmission for the
    two N\'eel-spin sectors (red/blue) at $h_z = 0$ and $0.3$ (dashed/solid). (b) Current and
    (c) differential conductance for different values of $h_z$. (d) Current
    polarization, $P^z = (I_+ - I_-)/I$. In all cases, $\theta = 0$ and $\beta_d = 0.15$.
  }
  \label{fig:3}
\end{figure}

\paragraph{Altermagnetic quantum dots.}

We consider a low-carrier-density altermagnet where a QD can be electrostatically defined,
Fig.~\ref{fig:1}, described by the Hamiltonian
\begin{equation}
  \label{eq:full-hamiltonian}
  \begin{split}
    \mathcal H & = \varepsilon_g + \frac{k_x^2 + k_y^2}{2m} + \frac{1}{2m} \left( \frac{x^2}{\ell_x^4} + \frac{y^2}{\ell_y^4} \right)              \\
               & \quad + \beta_d\left( k_x^2 - k_y^2 \right)\sigma_3 + \frac{h_z}{2}\,\sigma_3 + \alpha\left( \sigma_2 k_x - \sigma_1 k_y \right),
  \end{split}
\end{equation}
where $\varepsilon_g$ is the QD level energy that can be controlled through a
plunger gate, $m$ and $(k_x, k_y)$ are the carrier effective mass and in-plane momentum,
$\beta_d$ parametrizes the altermagnetic $d$-wave dispersion, $h_z$ is the
Zeeman-induced spin splitting along the N\'eel vector $\hat{\bm n}$, which we take perpendicular to
the plane, and $\bm{\sigma} = (\sigma_1,\sigma_2,\sigma_3)$ are Pauli matrices with $\sigma_3$ along
$\hat{\bm n}$. We set $\hbar = e = 1$ throughout. We use parabolic confinement with characteristic
lengths $\ell_x$ and $\ell_y$, although the spin selectivity does not depend on the specific form of
the confining potential, see the Supplemental Material (SM)~\cite{SM}. The last term in
Eq.~\eqref{eq:full-hamiltonian} is Rashba spin--orbit coupling with strength $\alpha$, which
mixes the two spin sectors. We set $\alpha = 0$ in the first part of the Letter, as it does not
significantly affect the presented results, and discuss its role in the last part. In this case, the
spin projection on the N\'eel vector ($\sigma_3$) is conserved, $\mathcal{H}$ block-diagonalizes
into two spin sectors $s = \pm1$, and the altermagnetic term $s\beta_d(k_x^2 - k_y^2)$ produces
opposite elliptic distortions of the confinement potential for the two spins as the two spins
acquire exchanged effective masses along $x$ and $y$ and, hence, oppositely squeezed orbitals.

For each of the spins, Eq.~\eqref{eq:full-hamiltonian} describes an anisotropic
oscillator with frequencies $\Omega_{x,s} = \omega_x\sqrt{1 + 2sb}$ and
$\Omega_{y,s} = \omega_y\sqrt{1 - 2sb}$ where $\omega_{x,y} = 1 / (m\ell_{x,y}^2)$
are the bare confinement frequencies, with $\ell_x = \ell_y \equiv \ell$ and
$\omega_x = \omega_y \equiv \omega_0$ for a circular dot, and $b \equiv m\beta_d$ is
the mass anisotropy of each spin band such that
$m_{y,s}/m_{x,s} = (1 + 2sb)/(1 - 2sb)$ for $|b| < 1/2$~\footnote{Stability of this
quadratic continuum model requires $|b| < 1/2$: at the boundary one kinetic
coefficient vanishes, and beyond it the spectrum is unbounded below.}.

The spin-dependent confinement potential leads to a spin splitting that depends on
the orbital indices, $n_{x,y}$, as
\begin{equation}
  \label{eq:level-spin-splitting}
  \begin{split}
    \Delta_{n_x,n_y} & = h_z + D\left[ \omega_x\left( n_x + \frac{1}{2} \right) - \omega_y\left( n_y + \frac{1}{2} \right) \right], \\
    D                & = \sqrt{1 + 2b} - \sqrt{1 - 2b} = 2b + O(b^3) ,
  \end{split}
\end{equation}
where $\Delta_{n_x,n_y} = E_{n_x,n_y,+} - E_{n_x,n_y,-}$. For a circular dot, $D\omega_0(n_x - n_y)$
vanishes for $n_x = n_y$, so the ground doublet stays degenerate, while levels with
$n_x \neq n_y$ split by an amount that grows in the higher shells [Fig.~\ref{fig:2}(a)]. In
contrast, breaking the $C_4$ symmetry by, e.g., making elongated QDs along the $x$ or $y$
directions, also splits the levels with $n_x = n_y$, including the ground doublet. This
property makes spin splitting dependent on the confinement anisotropy and,
therefore, gate tunable [Fig.~\ref{fig:2}(b)].

In the following, we focus on the lowest-energy doublet with $(n_x,n_y) = (0,0)$, whose energy
we denote by $\varepsilon_s = \varepsilon_g + s\Delta/2$, with
$\Delta \equiv \Delta_{0,0}$. The different effective masses of the two spin species
along the $x$ and $y$ axes and the exponential decay of their wavefunctions make the
results qualitatively similar for higher-energy levels as well~\cite{SM}.

\paragraph{Spin-selective hybridization.}

We place tunnel contacts at polar angles $\theta_j$ and distance $R_{\mathrm{lead}}$ from the
dot center, $j = L,R$~\cite{SM}. Projected onto the lowest doublet, the hybridization of
contact $j$ is a positive-semidefinite Hermitian $2 \times 2$ matrix, given by
\begin{equation}
  \label{eq:effective-polarization}
  \Gamma_j
  = \bar{\Gamma}_j\left( \bm{1} + \bm{\Pi}_j \cdot \bm{\sigma} \right)
  = \bar{\Gamma}_j\left( \bm{1} + \Pi^z_j\sigma_3 + \bm\Pi^\perp_j \cdot \bm\sigma_\perp \right),
\end{equation}
with spin-averaged coupling $\bar\Gamma_j = \tr\Gamma_j/2$ and polarization
$\bm\Pi_j$; positivity requires $|\bm\Pi_j| \le 1$. We separate the component $\Pi^z_j$ along the
N\'eel axis from the transverse one, $\bm\Pi^\perp_j = (\Pi^x_j, \Pi^y_j)$, with
$\bm\sigma_\perp = (\sigma_1, \sigma_2)$. A contact pair lies on the same axis for
$\theta_R = \theta_L + \pi$ and is orthogonal for $\theta_R = \theta_L \pm \pi/2$. Spin
collinear or non-collinear configurations depend on the relative orientation of
$\bm\Pi_L$ and $\bm\Pi_R$. We consider same-axis pairs throughout; orthogonal pairs are analyzed
separately in the SM~\cite{SM}. Without spin--orbit coupling, $\sigma_3$ is conserved, so
$\bm\Pi^\perp_j = 0$, and the two N\'eel spins couple with rates
\begin{equation}
  \label{eq:harmonic-rates}
  \Gamma_{j, \pm} = \Gamma_{0j}e^{\pm A_j}, \qquad \Pi^z_j = \tanh A_j.
\end{equation}
For a parabolic QD and a narrow contact of angular half-width $\varphi$,
$A_j = b\rho^2\cos2\theta_j$ to leading order in $b$, with $\rho \equiv R_{\mathrm{lead}}/\ell$ and
$\Gamma_{0j} \propto \varphi\,e^{-\rho^2}$. Therefore, increasing the distance between the
leads and the QD decreases the tunnel rate, while increasing the spin polarization $|\Pi^z_j|$
toward unity.

Altermagnetism thus generates spin-dependent tunnel rates: the ratio
$\Gamma_{j,+}/\Gamma_{j,-} = e^{2A_j}$ grows exponentially with $b$
[Fig.~\ref{fig:2}(c)] and follows the $d$-wave form factor $\cos2\theta_j$
[Fig.~\ref{fig:2}(d)].

\paragraph{Transport characteristics.}

The spin-dependent rates shape transport through the dot. For a same-axis pair along the $x$ axis,
identical contacts give a transmission peak of unit height for each spin sector, and widths set by
$\Gamma_s \equiv \Gamma_{L,s} + \Gamma_{R,s}$, Fig.~\ref{fig:3}(a). A spin splitting
$\Delta$ separates the two peaks, and the current shows two steps of unequal height
[Fig.~\ref{fig:3}(b)], whose position in voltage is controlled by $\Delta$ and which appear in the
differential conductance, $G = \dd I/\dd V$, as peaks of unequal height [Figs.~\ref{fig:3}(c)]
for $k_{\mathrm B}T \gg \Gamma_s$~\footnote{For $k_{\mathrm B}T \ll \Gamma_s$, the two conductance
peaks have the same height but different width, as the transmission in Fig.~\ref{fig:3}(a).}
For $A_L = A_R \equiv A$, the height of the conductance peaks is
\begin{equation}
  \label{eq:peak-height}
  G^{\mathrm{max}}_s
  \sim    \frac{e^2}{h}\, \frac{\Gamma_{L,s}\Gamma_{R,s}}{\Gamma_s}\, \frac{1}{k_{\mathrm B}T}
  \propto e^{sA}.
\end{equation}
The resulting peak asymmetry is
\begin{equation}
  \label{eq:peak-asymmetry}
  \mathcal{A}
  \equiv  \frac{G^{\mathrm{max}}_{+} - G^{\mathrm{max}}_{-}}{G^{\mathrm{max}}_{+} + G^{\mathrm{max}}_{-}}
  \approx \Pi^z \qquad \left( k_{\mathrm B}T \gg \Gamma_s \right).
\end{equation}
The spin asymmetry is largest for contacts on the same principal axis ($x$ or $y$), vanishes
at the altermagnet nodes ($\pm \pi/4$, $\pm3\pi/4$ for a $d$-wave altermagnet), and is suppressed
by angular averaging~\cite{SM}.

The current through the QD is highly spin-polarized, as illustrated by the normalized
spin current, $P^z = (I_+ - I_-)/(I_+ + I_-)$, with $I_\pm$ being the current for the
two spin projections along the N\'eel axis, Fig.~\ref{fig:3}(d). This property persists
in the presence of spin--orbit coupling, which tilts the angle of the spin current
between the source and drain leads, see the SM~\cite{SM}.

\paragraph{Coulomb blockade.}

\begin{figure}[h]
  \centering
  \includegraphics[width=1.0\columnwidth]{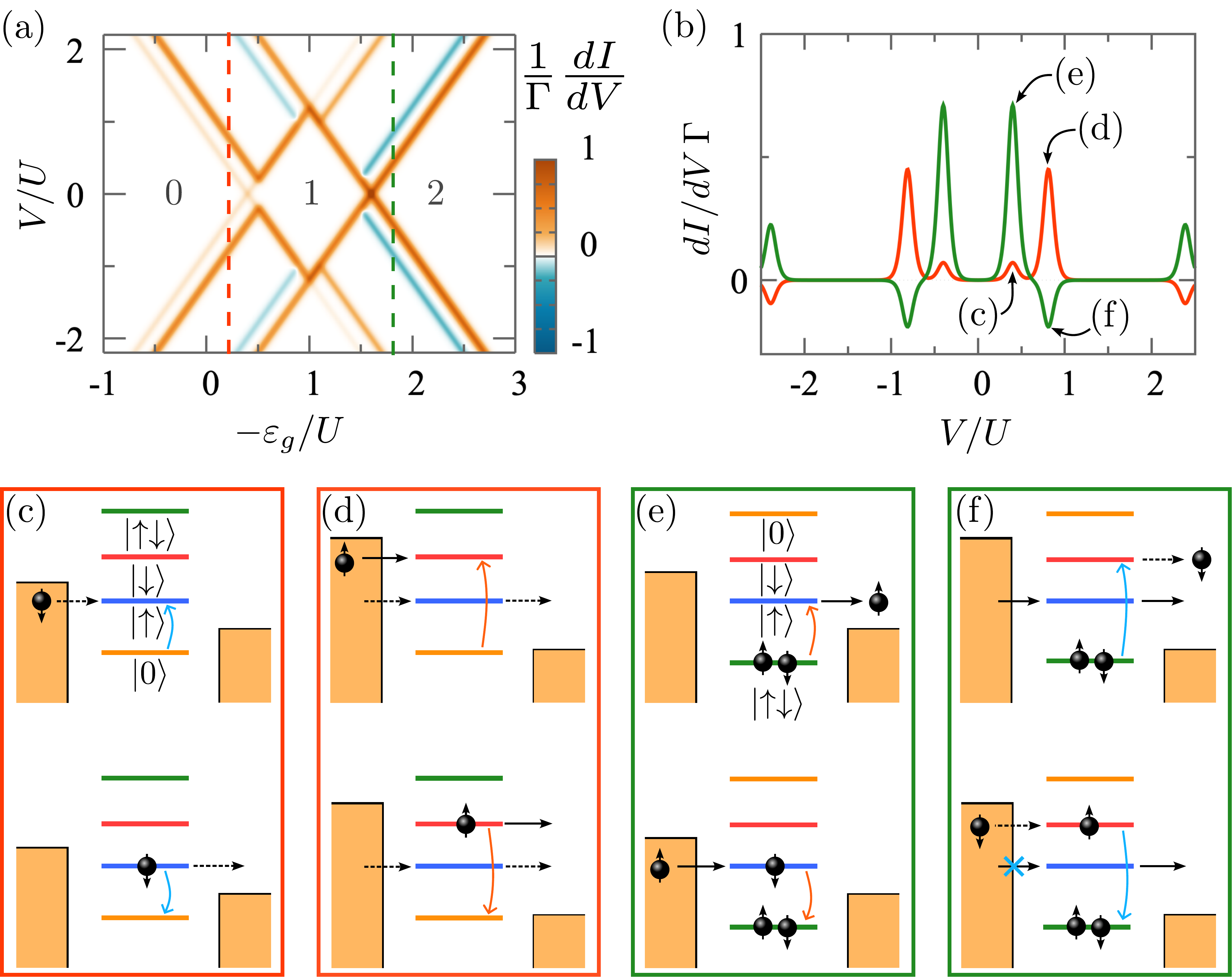}
  \caption{
    (a) Differential conductance $G = \dd I / \dd V$, normalized by the largest peak
    $G_{\max}$ of the cuts in (b), through the interacting dot for the rate ratio
    $\Gamma_{j,-} / \Gamma_{j,+} = 0.1$. (b) Conductance cuts at the gate settings
    $\varepsilon_g / U = 0.25$ (orange) and $-1.25$ (green). (c--f) sketch the dominant sequential
    processes at the marked features. Parameters are $\Delta = 0.2\,U$,
    $k_{\mathrm B}T = 0.02\,U$, and
    $\tr\Gamma_j = \Gamma_{j,+} + \Gamma_{j,-} = 0.002\,U$ for each lead,
    $\Gamma = \tr(\Gamma_L + \Gamma_R)$, with $\mu_{L,R} = \pm V/2$ and no spin mixing.
  }
  \label{fig:4}
\end{figure}

Coulomb repulsion in QDs fixes the number of electrons in the regime where the charging
energy, $U$, exceeds temperature and tunnel broadening. Restricted to the lowest orbital,
the dot is described by the Anderson model
\begin{equation}
  \label{eq:interacting-hamiltonian}
  H_{\mathrm{D}} = \sum_{s= \pm 1}\varepsilon_s\hat{n}_s + U\hat{n}_+\hat{n}_-,
\end{equation}
where $\hat n_s = d_s^\dagger d_s$. For $\bar\Gamma \ll k_{\mathrm B}T \ll U$, we use master
equations~\cite{Kirsanskas_2017_QmeQ, Kirsanskas_2018_Phenomenological,
Nathan_2020_Universal, Nathan_2021_Erratum}, to second order in the tunneling amplitudes
\begin{equation}
  \label{eq:master-equation}
  \dot\varrho = -i \left[ H_{\mathrm D} + H_{\mathrm{LS}},\varrho \right] + \mathcal D[\varrho],
\end{equation}
where $\varrho$ is the reduced density matrix of the dot and $\mathcal D$ describes
sequential transitions between the empty, singly, and doubly occupied dot states. These transitions
depend on the spin selectivity of the contacts: without spin mixing, an electron of spin $s$ tunnels
into the dot from lead $j$ at rate $\Gamma_{j,s}f_j(E)$ and out of it at rate
$\Gamma_{j,s}[1 - f_j(E)]$, with the spin-dependent $\Gamma_{j,s}$,
Eq.~\eqref{eq:harmonic-rates}. Here $f_j$ is the Fermi function of lead $j$, with chemical potential
$\mu_j$, and $E$ is the addition energy: $\varepsilon_s$ for an empty dot and $\varepsilon_s + U$
when spin $-s$ is already in the QD. The net charge transfer during these transitions gives the
current. The principal value part of the same tunneling expansion gives the Lamb shift
$H_{\mathrm{LS}}$ from virtual transitions~\cite{SM}.

The differential conductance features Coulomb diamonds, with well-defined electron
number at low bias, except close to the charge-degeneracy points,
Fig.~\ref{fig:4}(a). Near the $0\! \leftrightarrow \!1$ transition [orange cut in
Fig.~\ref{fig:4}(b)], the dominant process is the tunneling of a spin-down electron, leading to a
weak conductance peak, Fig.~\ref{fig:4}(c). Once the bias is increased to reach the strongly
coupled up level, a parallel channel opens, leading to a large conductance peak
[Fig.~\ref{fig:4}(d)].

Near the $1\! \leftrightarrow \!2$ transition (green cut) the roles reverse. The low-bias cycle
exchanges up electrons through the strongly coupled channel [Fig.~\ref{fig:4}(e)]. Above the
spin-down threshold, the dot can drop into the singly occupied up state, which it leaves only by
loading a down electron, the weakly tunnel-coupled spin. As the dynamics of the down spins is
slower, the dot is trapped there and the up current is blocked [Fig.~\ref{fig:4}(f)],
giving negative differential conductance. This feature is absent for equal rates for
the two spins and is therefore a direct signature of the spin-selective tunneling
originating from the altermagnetic band structure of the host material.

\paragraph{Exchange field from unpolarized leads.}

\begin{figure}[t]
  \centering
  \includegraphics[width=\columnwidth]{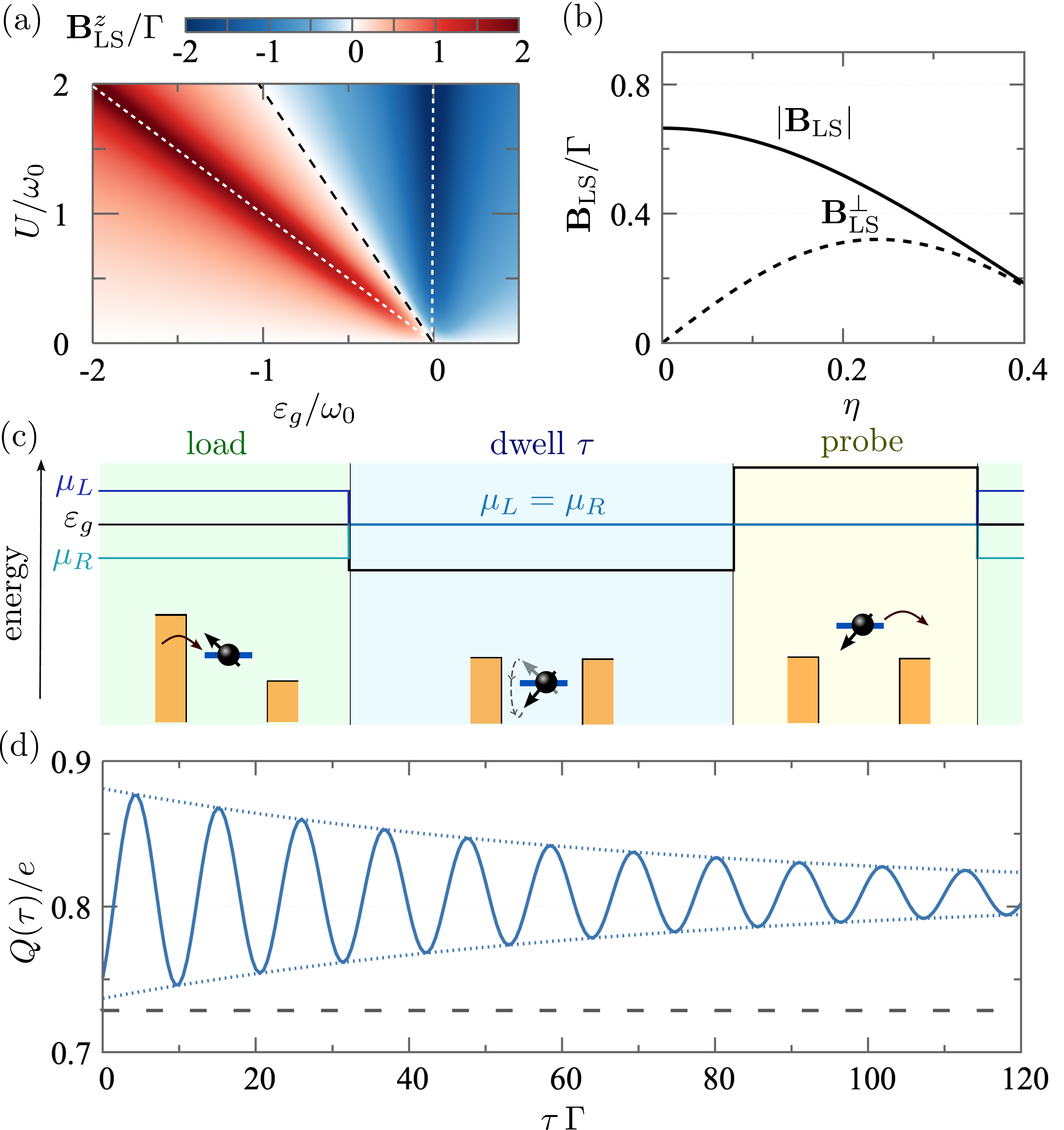}
  \caption{
    Lamb-shift-induced exchange field $\bm B_{\mathrm{LS}}$ on the singly occupied doublet and
    its readout. (a) $B^z_{\mathrm{LS}}$ at $\eta = 0$ against gate level and
    charging energy: it changes sign at $\varepsilon_g = -U/2$ (black dashed) and is largest near
    the charge-degeneracy lines $\varepsilon_g = 0$ and $\varepsilon_g = -U$ (white dotted). (b)
    $|\bm B_{\mathrm{LS}}|$ (solid) and its component transverse to the source polarization
    (dashed) against the Rashba strength $\eta$. (c) Stroboscopic sequence: an
    electron is loaded from the source; during the dwell $\tau$, at zero bias in the
    Coulomb valley, the pseudospin precesses; the probe lifts the level above both chemical
    potentials, and the dot empties into either lead. (d) Charge $Q_R(\tau)$ transferred to the
    drain per cycle, $I(\tau) = f_{\mathrm{rep}}Q_R(\tau)$, with (solid) and without
    (dashed) $\bm B_{\mathrm{LS}}$; dotted, the envelope $\propto e^{-\gamma\tau}$. Parameters:
    $b = 0.15$, $\rho = 3$, $\varphi = 0.15$, $\eta = 0.15$, $\bar\Gamma/U = 0.01$,
    $k_{\mathrm B}T/U = 0.05$, dwell at $\varepsilon_g/U = -0.25$, load bias $V/U = 0.4$.
  }
  \label{fig:5}
\end{figure}

In an altermagnetic QD, the Lamb shift $H_{\mathrm{LS}}$ due to the coupling to the
leads is spin dependent and splits the doublet. For a degenerate doublet
($\Delta = 0$), the effect is described by the effective Hamiltonian
\begin{equation}
  \label{eq:lamb-shift}
  H_{\mathrm{eff}} = (\varepsilon_g + \delta\varepsilon)\bm{1}
                     + \frac12\bm B_{\mathrm{LS}} \cdot \bm\sigma.
\end{equation}
The scalar shift ($\delta\varepsilon$) changes the addition energy, while the exchange field induced
by tunneling to the leads splits the doublet by $|\bm B_{\mathrm{LS}}|$, given by~\cite{SM}
\begin{align}
  \label{eq:exchange-field}
  \bm B_{\mathrm{LS}} & = \frac{1}{\pi}\sum_j F_j\,\bar{\Gamma}_j\,\bm{\Pi}_j,                                        \\
  F_j                 & \simeq \ln\frac{|\varepsilon_g - \mu_j|}{|\varepsilon_g + U - \mu_j|}, \label{eq:lead-factor}
\end{align}
where the lead factor $F_j$ is defined in the SM~\cite{SM} and the approximation holds for
$|\varepsilon_g - \mu_j|, |\varepsilon_g + U - \mu_j| \gg k_{\mathrm B}T$.

Unlike in QDs coupled to ferromagnetic leads, spin polarization in altermagnetic QDs
originates in the dot and requires no magnetism in the leads
[Eq.~\eqref{eq:harmonic-rates}], being tunable from the outside as the gate-defined confinement
sets $\bm\Pi_j$. The angle between the spin polarizations of the incoming and outgoing electrons is
determined by the angle between the contacts~\cite{SM} and the spin--orbit coupling, which is
tunable through electrostatic gates~\cite{Liu_2022_Gate}.

Without spin--orbit coupling, $\bm\Pi_j$ and $\bm B_{\mathrm{LS}}$ align with the
N\'eel axis, so the Lamb shift induces no torque on the electron spin that tunnels from the source.
For $\eta \equiv m\alpha\ell = \ell/\ell_{\mathrm{SO}} \ll 1$, spin--orbit coupling rotates the spin
of the tunneling electron. A contact at distance $R_{\mathrm{lead}}$ samples the orbital tail, where
the spin texture is rotated by an angle
$R_{\mathrm{lead}}/\ell_{\mathrm{SO}} = \eta\rho$. The hybridization therefore
acquires a transverse component $\bm\Pi^\perp_j$ in Eq.~\eqref{eq:effective-polarization},
linear in both $b$ and $\eta$, Fig.~\ref{fig:5}(b). The tilt on the spin is opposite at the
two contacts, leading to non-collinear spin-polarized currents in the source and
drain, while shifting the dot levels only at $O(\eta^2)$~\cite{SM}. For the
degenerate doublet with charging energy, the non-collinear polarizations also produce
negative differential conductance already in sequential tunneling, as in Fig.~\ref{fig:4}~\cite{SM}.

The spin precession can be read out with a stroboscopic sequence,
Fig.~\ref{fig:5}(c)~\cite{Hell_2015_Spin}. A load pulse lowers the QD level into the
bias window, so that only the source fills it, preparing the QD in a state with one electron,
$\varrho_1(0) \simeq \Gamma_L/\tr\Gamma_L$, whose pseudospin points along the source polarization
$\bm\Pi_L$. The plunger gate then moves the dot into the Coulomb valley, where sequential
tunneling is suppressed, and the electron spin can precess. The bias is removed during this
part. After a dwell time $\tau$, a probe pulse lifts the level above both chemical potentials, so
the dot empties into either lead. The dc current
$I(\tau) = f_{\mathrm{rep}}Q_R(\tau)$ of a sequence repeated at rate
$f_{\mathrm{rep}}$ records the charge $Q_R$ delivered to the drain per
cycle~\cite{Nakamura_1999_Coherent, Hayashi_2003_Coherent},
\begin{equation}
  \label{eq:pump-probe}
  Q_R(\tau) = Q_{\mathrm{bg}} + Q_1 e^{-\gamma\tau} \cos\!\left( \omega\tau + \Phi \right).
\end{equation}
Here $\omega$ is the precession frequency of the pseudospin and $\gamma$ its damping rate, given by
the complex-conjugate pair of eigenvalues $-\gamma \pm i\omega$ of the generator of
Eq.~\eqref{eq:master-equation} during the dwell. The amplitude $Q_1$ and phase
$\Phi$ depend on the load and probe pulses, and
$Q_{\mathrm{bg}} = Q_R(\tau \to \infty)$ includes the charge transferred during the
load~\cite{SM}. For $\bm B_{\mathrm{LS}} = 0$, the spin does not precess and $Q_R$ does not
depend on $\tau$ [Fig.~\ref{fig:5}(d), dashed]. For ideal source loading and a long probe,
with the field acting only during the dwell, identical same-axis contacts give
$Q_1 \simeq e|\bm\Pi_L^\perp|^2/2$ and $\Phi \simeq \pi$ in the weak-damping
limit~\cite{SM}. The oscillation amplitude therefore measures the transverse contact
polarization, while its frequency measures the exchange field. At first order,
\begin{equation}
  \label{eq:first-order-damping}
  \begin{split}
    \omega^{(1)} =      |\bm B_{\mathrm{LS}}|,\quad \gamma^{(1)}
                 \simeq \sum_j\bar\Gamma_j\,e^{-(\mu-\varepsilon_g)/k_{\mathrm B}T},
  \end{split}
\end{equation}
valid for $\mu - U/2 < \varepsilon_g < \mu$, with $\mu = \mu_L = \mu_R$ the common chemical
potential during the zero-bias dwell. The oscillation obtained by integrating
Eq.~\eqref{eq:master-equation} is therefore underdamped [Fig.~\ref{fig:5}(d)], with a
quality factor $\omega^{(1)}/\gamma^{(1)}$ that grows exponentially with the depth of the
dwell. For identical same-axis contacts at zero bias, the transverse exchange-field
components cancel. A small bare splitting along the N\'eel axis then gives
$\omega^{(1)} \simeq |\Delta + B^z_{\mathrm{LS}}|$, neglecting its effect on the
contact matrices and tunneling kernel. The readout requires
$\omega \lesssim \bar\Gamma$, since faster precession during the load and the exit
averages out the transverse pseudospin. The exchange field meets this automatically,
$|\bm B_{\mathrm{LS}}| \sim \bar\Gamma$ [Eq.~\eqref{eq:exchange-field}], whereas $\Delta$ must be
tuned. Cotunneling damps the oscillation further but leaves it
underdamped~\cite{SM}, and in the dc limit the field leaves a narrow dip in the valley conductance
for $\varepsilon_g = -U/2$~\cite{SM}.

\paragraph{Conclusion.}

Quantum confinement converts the momentum-dependent spin splitting of an altermagnet into distinct
spatial profiles for the two spin species. In quantum dots (QDs), this property induces a spin splitting
that is level-dependent and can be tuned by the electrostatic gates that set the confinement.
The different spatial profiles induce spin-selective tunneling to external leads, even if they are
spin-unpolarized, which results in current steps of unequal height, from which the spin asymmetry in
tunneling can be measured. Current through the system is strongly spin-polarized, with a
polarization that is gate-tunable. This picture is not affected by spin--orbit coupling or charging
energy. Weak spin--orbit coupling preserves spin selectivity, tilting the current spin
polarization axis between the source and drain leads. Charging energy makes it possible to measure
this tilting through a stroboscopic sequence that converts the spin precession into a dc current.
These findings establish quantum confinement as a route for directly probing
altermagnetic properties at the nanoscale.

Altermagnetic semiconductors, such as $\alpha$-MnTe~\cite{Krempasky_2024_Altermagnetic,
Lee_2024_Broken, Osumi_2024_Observation} and FeS~\cite{Takagi_2025_Spontaneous} in
films with the $c$ axis in the plane, are promising candidates for realizing QDs.
Their low carrier densities enable fast electrostatic control, desirable for spintronic and
quantum information devices, including altermagnet-based qubits~\cite{Abadillouriel_2026_All,
Vakili_2026_Gate, Tjernshaugen_2026_Superconducting}. The key ingredients have experimental
precedents: contact-induced exchange fields have been resolved in ferromagnet-contacted
dots~\cite{Hauptmann_2008_Electric, Hofstetter_2010_Ferromagnetic}, and pulsed-gate protocols
such as that of Fig.~\ref{fig:5}(c) are routine in spin qubits~\cite{Elzerman_2004_Single}.
The proposal also extends to nanoscale altermagnetic islands and impurity-induced surface
states~\cite{Sierda_2023_Quantum}, where spin-selective scanning tunneling microscopy can
visualize the anisotropic spin confinement~\cite{Yang_2026_Visualizing}.
Gate-defined altermagnetic dots thus offer a route to generating, manipulating, and electrically
reading out spin polarization using unpolarized electrodes and no applied magnetic field.

\paragraph{Data availability.}

The data that support the findings of this article, and the code that generates them and every
figure, are openly available~\cite{Maiani_2026_Code}.

\paragraph{Acknowledgments.}

Work supported by the Spanish Ministry of Science, Innovation, and Universities through
Grants RYC2022-037527-I, PID2023-148257NA-I00, PID2024-161156NB-I00, and PID2025-169814NB-I00
funded by MICIU/\allowbreak AEI/\allowbreak 10.13039/\allowbreak 501100011033 and by
FSE+ and FEDER, UE, ``ERDF A way of making Europe'' and European Union Next
Generation EU/PRTR. JCAU, RSS, and RA acknowledge the support of the CSIC's Quantum
Technologies Platform (QTEP) and the Severo Ochoa Centers of Excellence program through
Grant CEX2024-001445-S. RSS acknowledges funding from the Spanish Comunidad de Madrid (CM)
``Talento Program'' (Project No. 2022-T1/IND-24070). AM acknowledges funding from
the Wallenberg Initiative on Networks and Quantum Information (WINQ).

\paragraph{AI use disclosure.}

Claude Code (Anthropic; Claude Opus 4.8--5.5) and Codex (OpenAI; GPT-5.5--6.1) assisted, under the
authors' direction, with analytic derivation extension, numerical and plotting code. The
authors verified the derivations numerically and reviewed all code and output.
\bibliography{amdot.bib}

@article{Kouwenhoven_1991_Quantized,
  author    = {Kouwenhoven, L. P. and Johnson, A. T. and van der Vaart, N. C. and Harmans, C. J. P. M. and Foxon, C. T.},
  title     = {Quantized current in a quantum-dot turnstile using oscillating tunnel barriers},
  journal   = {Phys. Rev. Lett.},
  year      = {1991},
  volume    = {67},
  number    = {12},
  pages     = {1626--1629},
  month     = {Sept},
  doi       = {10.1103/PhysRevLett.67.1626},
  url       = {http://dx.doi.org/10.1103/PhysRevLett.67.1626},
  issn      = {0031-9007},
  publisher = {American Physical Society (APS)},
}

@article{Meservey_1994_Spin,
  author    = {Meservey, R. and Tedrow, P. M.},
  title     = {Spin-polarized electron tunneling},
  journal   = {Phys. Rep.},
  year      = {1994},
  volume    = {238},
  number    = {4},
  pages     = {173--243},
  month     = {March},
  doi       = {10.1016/0370-1573(94)90105-8},
  url       = {http://dx.doi.org/10.1016/0370-1573(94)90105-8},
  issn      = {0370-1573},
  publisher = {Elsevier BV},
}

@article{Schoeller_1994_Mesoscopic,
  author    = {Schoeller, Herbert and Schön, Gerd},
  title     = {Mesoscopic quantum transport: Resonant tunneling in the presence of a strong Coulomb interaction},
  journal   = {Phys. Rev. B},
  year      = {1994},
  volume    = {50},
  number    = {24},
  pages     = {18436--18452},
  month     = {December},
  doi       = {10.1103/PhysRevB.50.18436},
  url       = {http://dx.doi.org/10.1103/PhysRevB.50.18436},
  issn      = {1095-3795},
  publisher = {American Physical Society (APS)},
}

@article{Loss_1998_Quantum,
  author    = {Loss, Daniel and DiVincenzo, David P.},
  title     = {Quantum computation with quantum dots},
  journal   = {Phys. Rev. A},
  year      = {1998},
  volume    = {57},
  number    = {1},
  pages     = {120--126},
  month     = {January},
  doi       = {10.1103/PhysRevA.57.120},
  url       = {http://dx.doi.org/10.1103/PhysRevA.57.120},
  issn      = {1094-1622},
  publisher = {American Physical Society (APS)},
}

@article{Nakamura_1999_Coherent,
  author    = {Nakamura, Y. and Pashkin, Yu. A. and Tsai, J. S.},
  title     = {Coherent control of macroscopic quantum states in a single-Cooper-pair box},
  journal   = {Nature},
  year      = {1999},
  volume    = {398},
  number    = {6730},
  pages     = {786--788},
  month     = {April},
  doi       = {10.1038/19718},
  url       = {http://dx.doi.org/10.1038/19718},
  issn      = {1476-4687},
  publisher = {Springer Science and Business Media LLC},
}

@article{Recher_2000_Quantum,
  author    = {Recher, Patrik and Sukhorukov, Eugene V. and Loss, Daniel},
  title     = {Quantum Dot as Spin Filter and Spin Memory},
  journal   = {Phys. Rev. Lett.},
  year      = {2000},
  volume    = {85},
  number    = {9},
  pages     = {1962--1965},
  month     = {August},
  doi       = {10.1103/PhysRevLett.85.1962},
  url       = {http://dx.doi.org/10.1103/PhysRevLett.85.1962},
  issn      = {1079-7114},
  publisher = {American Physical Society (APS)},
}

@article{Reimann_2002_Electronic,
  author    = {Reimann, Stephanie M. and Manninen, Matti},
  title     = {Electronic structure of quantum dots},
  journal   = {Rev. Mod. Phys.},
  year      = {2002},
  volume    = {74},
  number    = {4},
  pages     = {1283--1342},
  month     = {November},
  doi       = {10.1103/RevModPhys.74.1283},
  url       = {http://dx.doi.org/10.1103/RevModPhys.74.1283},
  issn      = {1539-0756},
  publisher = {American Physical Society (APS)},
}

@article{Hayashi_2003_Coherent,
  author    = {Hayashi, T. and Fujisawa, T. and Cheong, H. D. and Jeong, Y. H. and Hirayama, Y.},
  title     = {Coherent Manipulation of Electronic States in a Double Quantum Dot},
  journal   = {Phys. Rev. Lett.},
  year      = {2003},
  volume    = {91},
  number    = {22},
  pages     = {226804},
  month     = {November},
  doi       = {10.1103/PhysRevLett.91.226804},
  url       = {http://dx.doi.org/10.1103/PhysRevLett.91.226804},
  issn      = {1079-7114},
  publisher = {American Physical Society (APS)},
}

@article{Konig_2003_Interaction,
  author    = {König, Jürgen and Martinek, Jan},
  title     = {Interaction-Driven Spin Precession in Quantum-Dot Spin Valves},
  journal   = {Phys. Rev. Lett.},
  year      = {2003},
  volume    = {90},
  number    = {16},
  pages     = {166602},
  month     = {April},
  doi       = {10.1103/PhysRevLett.90.166602},
  url       = {http://dx.doi.org/10.1103/PhysRevLett.90.166602},
  issn      = {1079-7114},
  publisher = {American Physical Society (APS)},
}

@article{Braun_2004_Theory,
  author    = {Braun, Matthias and König, Jürgen and Martinek, Jan},
  title     = {Theory of transport through quantum-dot spin valves in the weak-coupling regime},
  journal   = {Phys. Rev. B},
  year      = {2004},
  volume    = {70},
  number    = {19},
  pages     = {195345},
  month     = {November},
  doi       = {10.1103/PhysRevB.70.195345},
  url       = {http://dx.doi.org/10.1103/PhysRevB.70.195345},
  issn      = {1550-235X},
  publisher = {American Physical Society (APS)},
}

@article{Elzerman_2004_Single,
  author    = {Elzerman, J. M. and Hanson, R. and Willems van Beveren, L. H. and Witkamp, B. and Vandersypen, L. M. K. and Kouwenhoven, L. P.},
  title     = {Single-shot read-out of an individual electron spin in a quantum dot},
  journal   = {Nature},
  year      = {2004},
  volume    = {430},
  number    = {6998},
  pages     = {431--435},
  month     = {July},
  doi       = {10.1038/nature02693},
  url       = {http://dx.doi.org/10.1038/nature02693},
  issn      = {1476-4687},
  publisher = {Springer Science and Business Media LLC},
}

@article{Hanson_2004_Semiconductor,
  author    = {Hanson, R. and Vandersypen, L. M. K. and van Beveren, L. H. Willems and Elzerman, J. M. and Vink, I. T. and Kouwenhoven, L. P.},
  title     = {Semiconductor few-electron quantum dot operated as a bipolar spin filter},
  journal   = {Phys. Rev. B},
  year      = {2004},
  volume    = {70},
  number    = {24},
  pages     = {241304},
  month     = {December},
  doi       = {10.1103/PhysRevB.70.241304},
  url       = {http://dx.doi.org/10.1103/PhysRevB.70.241304},
  issn      = {1550-235X},
  publisher = {American Physical Society (APS)},
}

@article{Zutic_2004_Spintronics,
  author    = {Žutić, Igor and Fabian, Jaroslav and Das Sarma, S.},
  title     = {Spintronics: Fundamentals and applications},
  journal   = {Rev. Mod. Phys.},
  year      = {2004},
  volume    = {76},
  number    = {2},
  pages     = {323--410},
  month     = {April},
  doi       = {10.1103/RevModPhys.76.323},
  url       = {http://dx.doi.org/10.1103/RevModPhys.76.323},
  issn      = {1539-0756},
  publisher = {American Physical Society (APS)},
}

@article{Blumenthal_2007_Gigahertz,
  author    = {Blumenthal, M. D. and Kaestner, B. and Li, L. and Giblin, S. and Janssen, T. J. B. M. and Pepper, M. and Anderson, D. and Jones, G. and Ritchie, D. A.},
  title     = {Gigahertz quantized charge pumping},
  journal   = {Nat. Phys.},
  year      = {2007},
  volume    = {3},
  number    = {5},
  pages     = {343--347},
  month     = {April},
  doi       = {10.1038/nphys582},
  url       = {http://dx.doi.org/10.1038/nphys582},
  issn      = {1745-2481},
  publisher = {Springer Science and Business Media LLC},
}

@article{Hanson_2007_Spins,
  author    = {Hanson, R. and Kouwenhoven, L. P. and Petta, J. R. and Tarucha, S. and Vandersypen, L. M. K.},
  title     = {Spins in few-electron quantum dots},
  journal   = {Rev. Mod. Phys.},
  year      = {2007},
  volume    = {79},
  number    = {4},
  pages     = {1217--1265},
  month     = oct,
  doi       = {10.1103/revmodphys.79.1217},
  issn      = {1539-0756},
  publisher = {American Physical Society (APS)},
}

@article{Weymann_2007_Cotunneling,
  author    = {Weymann, Ireneusz and Barnaś, Józef},
  title     = {Cotunneling through quantum dots coupled to magnetic leads: Zero-bias anomaly for noncollinear magnetic configurations},
  journal   = {Phys. Rev. B},
  year      = {2007},
  volume    = {75},
  number    = {15},
  pages     = {155308},
  month     = {April},
  doi       = {10.1103/PhysRevB.75.155308},
  url       = {http://dx.doi.org/10.1103/PhysRevB.75.155308},
  issn      = {1550-235X},
  publisher = {American Physical Society (APS)},
}

@article{Fert_2008_Nobel,
  author    = {Fert, Albert},
  title     = {Nobel Lecture: Origin, development, and future of spintronics},
  journal   = {Rev. Mod. Phys.},
  year      = {2008},
  volume    = {80},
  number    = {4},
  pages     = {1517--1530},
  month     = {December},
  doi       = {10.1103/RevModPhys.80.1517},
  url       = {http://dx.doi.org/10.1103/RevModPhys.80.1517},
  issn      = {1539-0756},
  publisher = {American Physical Society (APS)},
}

@article{Hauptmann_2008_Electric,
  author    = {Hauptmann, J. R. and Paaske, J. and Lindelof, P. E.},
  title     = {Electric-field-controlled spin reversal in a quantum dot with ferromagnetic contacts},
  journal   = {Nat. Phys.},
  year      = {2008},
  volume    = {4},
  number    = {5},
  pages     = {373--376},
  month     = {March},
  doi       = {10.1038/nphys931},
  url       = {http://dx.doi.org/10.1038/nphys931},
  issn      = {1745-2481},
  publisher = {Springer Science and Business Media LLC},
}

@article{Leijnse_2008_Kinetic,
  author    = {Leijnse, M. and Wegewijs, M. R.},
  title     = {Kinetic equations for transport through single-molecule transistors},
  journal   = {Phys. Rev. B},
  year      = {2008},
  volume    = {78},
  number    = {23},
  pages     = {235424},
  month     = {December},
  doi       = {10.1103/PhysRevB.78.235424},
  url       = {http://dx.doi.org/10.1103/PhysRevB.78.235424},
  issn      = {1550-235X},
  publisher = {American Physical Society (APS)},
}

@article{Timm_2008_Tunneling,
  author    = {Timm, Carsten},
  title     = {Tunneling through molecules and quantum dots: Master-equation approaches},
  journal   = {Phys. Rev. B},
  year      = {2008},
  volume    = {77},
  number    = {19},
  pages     = {195416},
  month     = {May},
  doi       = {10.1103/PhysRevB.77.195416},
  url       = {http://dx.doi.org/10.1103/PhysRevB.77.195416},
  issn      = {1550-235X},
  publisher = {American Physical Society (APS)},
}

@article{Hofstetter_2010_Ferromagnetic,
  author    = {Hofstetter, L. and Geresdi, A. and Aagesen, M. and Nygård, J. and Schönenberger, C. and Csonka, S.},
  title     = {Ferromagnetic Proximity Effect in a Ferromagnet–Quantum-Dot–Superconductor Device},
  journal   = {Phys. Rev. Lett.},
  year      = {2010},
  volume    = {104},
  number    = {24},
  pages     = {246804},
  month     = {June},
  doi       = {10.1103/PhysRevLett.104.246804},
  url       = {http://dx.doi.org/10.1103/PhysRevLett.104.246804},
  issn      = {1079-7114},
  publisher = {American Physical Society (APS)},
}

@article{Koller_2010_Density,
  author  = {Koller, S. and Grifoni, M. and Leijnse, M. and Wegewijs, M. R.},
  title   = {Density-operator approaches to transport through interacting quantum dots: Simplifications in fourth-order perturbation theory},
  journal = {Phys. Rev. B},
  year    = {2010},
  volume  = {82},
  number  = {23},
  pages   = {235307},
  doi     = {10.1103/PhysRevB.82.235307},
  url     = {https://doi.org/10.1103/PhysRevB.82.235307},
}

@article{Timm_2011_Time,
  author    = {Timm, Carsten},
  title     = {Time-convolutionless master equation for quantum dots: Perturbative expansion to arbitrary order},
  journal   = {Phys. Rev. B},
  year      = {2011},
  volume    = {83},
  number    = {11},
  pages     = {115416},
  month     = {March},
  doi       = {10.1103/PhysRevB.83.115416},
  url       = {http://dx.doi.org/10.1103/PhysRevB.83.115416},
  issn      = {1550-235X},
  publisher = {American Physical Society (APS)},
}

@article{Pekola_2013_Single,
  author    = {Pekola, Jukka P. and Saira, Olli-Pentti and Maisi, Ville F. and Kemppinen, Antti and Möttönen, Mikko and Pashkin, Yuri A. and Averin, Dmitri V.},
  title     = {Single-electron current sources: Toward a refined definition of the ampere},
  journal   = {Rev. Mod. Phys.},
  year      = {2013},
  volume    = {85},
  number    = {4},
  pages     = {1421--1472},
  month     = {October},
  doi       = {10.1103/RevModPhys.85.1421},
  url       = {http://dx.doi.org/10.1103/RevModPhys.85.1421},
  issn      = {1539-0756},
  publisher = {American Physical Society (APS)},
}

@article{Hell_2015_Spin,
  author    = {Hell, M. and Sothmann, B. and Leijnse, M. and Wegewijs, M. R. and König, J.},
  title     = {Spin resonance without spin splitting},
  journal   = {Phys. Rev. B},
  year      = {2015},
  volume    = {91},
  number    = {19},
  pages     = {195404},
  month     = {May},
  doi       = {10.1103/PhysRevB.91.195404},
  url       = {http://dx.doi.org/10.1103/PhysRevB.91.195404},
  issn      = {1550-235X},
  publisher = {American Physical Society (APS)},
}

@article{Kirsanskas_2017_QmeQ,
  author    = {Kiršanskas, Gediminas and Pedersen, Jonas Nyvold and Karlström, Olov and Leijnse, Martin and Wacker, Andreas},
  title     = {{QmeQ} 1.0: An open-source Python package for calculations of transport through quantum dot devices},
  journal   = {Comput. Phys. Commun.},
  year      = {2017},
  volume    = {221},
  pages     = {317--342},
  month     = {December},
  doi       = {10.1016/j.cpc.2017.07.024},
  url       = {http://dx.doi.org/10.1016/j.cpc.2017.07.024},
  issn      = {0010-4655},
  publisher = {Elsevier BV},
}

@article{Baltz_2018_Antiferromagnetic,
  author    = {Baltz, V. and Manchon, A. and Tsoi, M. and Moriyama, T. and Ono, T. and Tserkovnyak, Y.},
  title     = {Antiferromagnetic spintronics},
  journal   = {Rev. Mod. Phys.},
  year      = {2018},
  volume    = {90},
  number    = {1},
  pages     = {015005},
  month     = {February},
  doi       = {10.1103/RevModPhys.90.015005},
  url       = {http://dx.doi.org/10.1103/RevModPhys.90.015005},
  issn      = {1539-0756},
  publisher = {American Physical Society (APS)},
}

@article{Kirsanskas_2018_Phenomenological,
  author    = {Kiršanskas, Gediminas and Franckié, Martin and Wacker, Andreas},
  title     = {Phenomenological position and energy resolving Lindblad approach to quantum kinetics},
  journal   = {Phys. Rev. B},
  year      = {2018},
  volume    = {97},
  number    = {3},
  pages     = {035432},
  month     = {January},
  doi       = {10.1103/PhysRevB.97.035432},
  url       = {http://dx.doi.org/10.1103/PhysRevB.97.035432},
  issn      = {2469-9969},
  publisher = {American Physical Society (APS)},
}

@article{Ahn_2019_Antiferromagnetism,
  author    = {Ahn, Kyo-Hoon and Hariki, Atsushi and Lee, Kwan-Woo and Kuneš, Jan},
  title     = {Antiferromagnetism in {RuO$_2$} as $d$-wave {Pomeranchuk} instability},
  journal   = {Phys. Rev. B},
  year      = {2019},
  volume    = {99},
  number    = {18},
  pages     = {184432},
  month     = may,
  doi       = {10.1103/physrevb.99.184432},
  issn      = {2469-9969},
  publisher = {American Physical Society (APS)},
}

@article{Hayami_2019_Momentum,
  author    = {Hayami, Satoru and Yanagi, Yuki and Kusunose, Hiroaki},
  title     = {{Momentum-Dependent Spin Splitting by Collinear Antiferromagnetic Ordering}},
  journal   = {J. Phys. Soc. Jpn.},
  year      = {2019},
  volume    = {88},
  number    = {12},
  pages     = {123702},
  month     = dec,
  doi       = {10.7566/jpsj.88.123702},
  issn      = {1347-4073},
  publisher = {Physical Society of Japan},
}

@article{Naka_2019_Spin,
  author    = {Naka, Makoto and Hayami, Satoru and Kusunose, Hiroaki and Yanagi, Yuki and Motome, Yukitoshi and Seo, Hitoshi},
  title     = {Spin current generation in organic antiferromagnets},
  journal   = {Nat. Commun.},
  year      = {2019},
  volume    = {10},
  number    = {1},
  pages     = {4305},
  month     = sep,
  doi       = {10.1038/s41467-019-12229-y},
  issn      = {2041-1723},
  publisher = {Springer Science and Business Media LLC},
}

@article{Naka_2020_Anomalous,
  author    = {Naka, Makoto and Hayami, Satoru and Kusunose, Hiroaki and Yanagi, Yuki and Motome, Yukitoshi and Seo, Hitoshi},
  title     = {{Anomalous Hall effect in $\kappa$-type organic antiferromagnets}},
  journal   = {Phys. Rev. B},
  year      = {2020},
  volume    = {102},
  number    = {7},
  pages     = {075112},
  month     = aug,
  doi       = {10.1103/physrevb.102.075112},
  issn      = {2469-9969},
  publisher = {American Physical Society (APS)},
}

@article{Nathan_2020_Universal,
  author    = {Nathan, Frederik and Rudner, Mark S.},
  title     = {Universal Lindblad equation for open quantum systems},
  journal   = {Phys. Rev. B},
  year      = {2020},
  volume    = {102},
  number    = {11},
  pages     = {115109},
  month     = {September},
  doi       = {10.1103/PhysRevB.102.115109},
  url       = {http://dx.doi.org/10.1103/PhysRevB.102.115109},
  issn      = {2469-9969},
  publisher = {American Physical Society (APS)},
}

@article{Gonzalezhernandez_2021_Efficient,
  author    = {González-Hernández, Rafael and Šmejkal, Libor and Výborný, Karel and Yahagi, Yuta and Sinova, Jairo and Jungwirth, Tomáš and Železný, Jakub},
  title     = {Efficient Electrical Spin Splitter Based on Nonrelativistic Collinear
               Antiferromagnetism},
  journal   = {Phys. Rev. Lett.},
  year      = {2021},
  volume    = {126},
  number    = {12},
  pages     = {127701},
  month     = {March},
  doi       = {10.1103/PhysRevLett.126.127701},
  url       = {http://dx.doi.org/10.1103/PhysRevLett.126.127701},
  issn      = {1079-7114},
  publisher = {American Physical Society (APS)},
}

@article{Nathan_2021_Erratum,
  author    = {Nathan, Frederik and Rudner, Mark S.},
  title     = {Erratum: Universal Lindblad equation for open quantum systems [Phys. Rev. B {\bf 102}, 115109 (2020)]},
  journal   = {Phys. Rev. B},
  year      = {2021},
  volume    = {104},
  number    = {11},
  pages     = {119901},
  month     = {September},
  doi       = {10.1103/PhysRevB.104.119901},
  url       = {http://dx.doi.org/10.1103/PhysRevB.104.119901},
  issn      = {2469-9969},
  publisher = {American Physical Society (APS)},
}

@article{Liu_2022_Gate,
  author    = {Liu, He and Zhang, Ting and Wang, Ke and Gao, Fei and Xu, Gang and Zhang, Xin and Li, Shu-Xiao and Cao, Gang and Wang, Ting and Zhang, Jianjun and Hu, Xuedong and Li, Hai-Ou and Guo, Guo-Ping},
  title     = {Gate-Tunable Spin-Orbit Coupling in a Germanium Hole Double Quantum Dot},
  journal   = {Phys. Rev. Appl.},
  year      = {2022},
  volume    = {17},
  pages     = {044052},
  numpages  = {9},
  month     = {Apr},
  doi       = {10.1103/PhysRevApplied.17.044052},
  url       = {https://link.aps.org/doi/10.1103/PhysRevApplied.17.044052},
  issue     = {4},
  publisher = {American Physical Society},
}

@article{Mazin_2022_Editorial,
  author    = {Mazin, Igor},
  title     = {{Editorial: Altermagnetism—A New Punch Line of Fundamental Magnetism}},
  journal   = {Phys. Rev. X},
  year      = {2022},
  volume    = {12},
  number    = {4},
  pages     = {040002},
  month     = dec,
  doi       = {10.1103/physrevx.12.040002},
  issn      = {2160-3308},
  publisher = {American Physical Society (APS)},
}

@article{Smejkal_2022_Giant,
  author    = {Šmejkal, Libor and Hellenes, Anna Birk and González-Hernández, Rafael and Sinova, Jairo and Jungwirth, Tomas},
  title     = {{Giant and Tunneling Magnetoresistance in Unconventional Collinear
  Antiferromagnets with Nonrelativistic Spin-Momentum Coupling}},
  journal   = {Phys. Rev. X},
  year      = {2022},
  volume    = {12},
  number    = {1},
  pages     = {011028},
  month     = feb,
  doi       = {10.1103/physrevx.12.011028},
  issn      = {2160-3308},
  publisher = {American Physical Society (APS)},
}

@article{Smejkal_2022_Emerging,
  author    = {Šmejkal, Libor and Sinova, Jairo and Jungwirth, Tomas},
  title     = {{Emerging Research Landscape of Altermagnetism}},
  journal   = {Phys. Rev. X},
  year      = {2022},
  volume    = {12},
  number    = {4},
  pages     = {040501},
  month     = dec,
  doi       = {10.1103/physrevx.12.040501},
  issn      = {2160-3308},
  publisher = {American Physical Society (APS)},
}

@article{Burkard_2023_Semiconductor,
  author    = {Burkard, Guido and Ladd, Thaddeus D. and Pan, Andrew and Nichol, John M. and Petta, Jason R.},
  title     = {Semiconductor spin qubits},
  journal   = {Rev. Mod. Phys.},
  year      = {2023},
  volume    = {95},
  number    = {2},
  pages     = {025003},
  month     = {June},
  doi       = {10.1103/RevModPhys.95.025003},
  url       = {http://dx.doi.org/10.1103/RevModPhys.95.025003},
  issn      = {1539-0756},
  publisher = {American Physical Society (APS)},
}

@article{Sierda_2023_Quantum,
  author    = {Sierda, E. and Huang, X. and Badrtdinov, D. I. and Kiraly, B. and Knol, E. J. and Groenenboom, G. C. and Katsnelson, M. I. and Rösner, M. and Wegner, D. and Khajetoorians, A. A.},
  title     = {Quantum simulator to emulate lower-dimensional molecular structure},
  journal   = {Science},
  year      = {2023},
  volume    = {380},
  number    = {6649},
  pages     = {1048--1052},
  month     = {June},
  doi       = {10.1126/science.adf2685},
  url       = {http://dx.doi.org/10.1126/science.adf2685},
  issn      = {1095-9203},
  publisher = {American Association for the Advancement of Science (AAAS)},
}

@article{Amin_2024_Nanoscale,
  author    = {Amin, O. J. and Dal Din, A. and Golias, E. and Niu, Y. and Zakharov, A. and Fromage, S. C. and Fields, C. J. B. and Heywood, S. L. and Cousins, R. B. and Maccherozzi, F. and Krempaský, J. and Dil, J. H. and Kriegner, D. and Kiraly, B. and Campion, R. P. and Rushforth, A. W. and Edmonds, K. W. and Dhesi, S. S. and Šmejkal, L. and Jungwirth, T. and Wadley, P.},
  title     = {Nanoscale imaging and control of altermagnetism in {MnTe}},
  journal   = {Nature},
  year      = {2024},
  volume    = {636},
  number    = {8042},
  pages     = {348--353},
  month     = {December},
  doi       = {10.1038/s41586-024-08234-x},
  url       = {http://dx.doi.org/10.1038/s41586-024-08234-x},
  issn      = {1476-4687},
  publisher = {Springer Science and Business Media LLC},
}

@article{Bai_2024_Altermagnetism,
  author    = {Bai, Ling and Feng, Wanxiang and Liu, Siyuan and Šmejkal, Libor and Mokrousov, Yuriy and Yao, Yugui},
  title     = {{Altermagnetism: Exploring New Frontiers in Magnetism and Spintronics}},
  journal   = {Adv. Funct. Mater.},
  year      = {2024},
  volume    = {34},
  number    = {49},
  pages     = {2409327},
  month     = sep,
  doi       = {10.1002/adfm.202409327},
  issn      = {1616-3028},
  publisher = {Wiley},
}

@article{Krempasky_2024_Altermagnetic,
  author    = {Krempaský, J. and Šmejkal, L. and D’Souza, S. W. and Hajlaoui, M. and Springholz, G. and Uhlířová, K. and Alarab, F. and Constantinou, P. C. and Strocov, V. and Usanov, D. and Pudelko, W. R. and González-Hernández, R. and Birk Hellenes, A. and Jansa, Z. and Reichlová, H. and Šobáň, Z. and Gonzalez Betancourt, R. D. and Wadley, P. and Sinova, J. and Kriegner, D. and Minár, J. and Dil, J. H. and Jungwirth, T.},
  title     = {Altermagnetic lifting of Kramers spin degeneracy},
  journal   = {Nature},
  year      = {2024},
  volume    = {626},
  number    = {7999},
  pages     = {517--522},
  month     = feb,
  doi       = {10.1038/s41586-023-06907-7},
  url       = {https://doi.org/10.1038/s41586-023-06907-7},
  issn      = {1476-4687},
  publisher = {Springer Science and Business Media LLC},
}

@article{Lee_2024_Broken,
  author    = {Lee, Suyoung and Lee, Sangjae and Jung, Saegyeol and Jung, Jiwon and Kim, Donghan and Lee, Yeonjae and Seok, Byeongjun and Kim, Jaeyoung and Park, Byeong Gyu and Šmejkal, Libor and Kang, Chang-Jong and Kim, Changyoung},
  title     = {Broken Kramers Degeneracy in Altermagnetic {MnTe}},
  journal   = {Phys. Rev. Lett.},
  year      = {2024},
  volume    = {132},
  number    = {3},
  pages     = {036702},
  month     = jan,
  doi       = {10.1103/PhysRevLett.132.036702},
  url       = {https://doi.org/10.1103/PhysRevLett.132.036702},
  issn      = {1079-7114},
  publisher = {American Physical Society},
}

@article{Osumi_2024_Observation,
  author    = {Osumi, T. and Souma, S. and Aoyama, T. and Yamauchi, K. and Honma, A. and Nakayama, K. and Takahashi, T. and Ohgushi, K. and Sato, T.},
  title     = {Observation of a giant band splitting in altermagnetic {MnTe}},
  journal   = {Phys. Rev. B},
  year      = {2024},
  volume    = {109},
  number    = {11},
  pages     = {115102},
  month     = {March},
  doi       = {10.1103/PhysRevB.109.115102},
  url       = {http://dx.doi.org/10.1103/PhysRevB.109.115102},
  issn      = {2469-9969},
  publisher = {American Physical Society (APS)},
}

@article{Song_2025_Altermagnets,
  author    = {Song, Cheng and Bai, Hua and Zhou, Zhiyuan and Han, Lei and Reichlova, Helena and Dil, J. Hugo and Liu, Junwei and Chen, Xianzhe and Pan, Feng},
  title     = {Altermagnets as a new class of functional materials},
  journal   = {Nat. Rev. Mater.},
  year      = {2025},
  volume    = {10},
  number    = {6},
  pages     = {473--485},
  month     = feb,
  doi       = {10.1038/s41578-025-00779-1},
  issn      = {2058-8437},
  publisher = {Springer Science and Business Media LLC},
}

@article{Takagi_2025_Spontaneous,
  author    = {Takagi, Rina and Hirakida, Ryosuke and Settai, Yuki and Oiwa, Rikuto and Takagi, Hirotaka and Kitaori, Aki and Yamauchi, Kensei and Inoue, Hiroki and Yamaura, Jun-ichi and Nishio-Hamane, Daisuke and Itoh, Shinichi and Aji, Seno and Saito, Hiraku and Nakajima, Taro and Nomoto, Takuya and Arita, Ryotaro and Seki, Shinichiro},
  title     = {Spontaneous Hall effect induced by collinear antiferromagnetic order at room temperature},
  journal   = {Nat. Mater.},
  year      = {2025},
  volume    = {24},
  number    = {1},
  pages     = {63--68},
  month     = {January},
  doi       = {10.1038/s41563-024-02058-w},
  url       = {http://dx.doi.org/10.1038/s41563-024-02058-w},
  issn      = {1476-4660},
  publisher = {Springer Science and Business Media LLC},
}

@article{Zarzuela_2025_Transport,
  author        = {Zarzuela, Ricardo and Jaeschke-Ubiergo, Rodrigo and Gomonay, Olena and Šmejkal, Libor and Sinova, Jairo},
  title         = {Transport theory and spin-transfer physics in $d$-wave altermagnets},
  journal       = {Phys. Rev. B},
  year          = {2025},
  volume        = {111},
  number        = {6},
  pages         = {064422},
  month         = feb,
  doi           = {10.1103/physrevb.111.064422},
  issn          = {2469-9969},
  archiveprefix = {arXiv},
  publisher     = {American Physical Society (APS)},
}

@misc{Abadillouriel_2026_All,
  author        = {Abadillo-Uriel, José Carlos and Maiani, Andrea and Cortijo, Alberto and Aguado, Ramón and Seoane Souto, Rubén},
  title         = {All-electrical dephasing-protected spin qubits in altermagnets},
  year          = {2026},
  month         = {jun},
  url           = {https://arxiv.org/abs/2606.26066},
  archiveprefix = {arXiv},
  eprint        = {2606.26066},
  primaryclass  = {cond-mat.mes-hall},
  updated       = {2026-06-24},
}

@article{Kirczenow_2026_Symmetry,
  author        = {Kirczenow, George},
  title         = {Symmetry, disorder and transport through altermagnetic quantum dots and their
               antiferromagnetic twins},
  journal       = {J. Phys.: Condens. Matter},
  year          = {2026},
  volume        = {38},
  number        = {17},
  pages         = {175302},
  month         = may,
  doi           = {10.1088/1361-648x/ae6431},
  eprint        = {2602.08154},
  archiveprefix = {arXiv},
}

@article{Liu_2025_Altermagnetic,
  author        = {Liu, Li-Shuo and Shao, Kai and Li, Hai-Dong and Wan, Xiangang and Chen, Wei and Xing, D. Y.},
  title         = {{Altermagnetic Spin Precession and Spin Transistor}},
  journal       = {Phys. Rev. Lett.},
  year          = {2026},
  volume        = {136},
  number        = {10},
  pages         = {106301},
  month         = mar,
  doi           = {10.1103/j3qj-77yj},
  date          = {2025-11-07},
  issn          = {1079-7114},
  archiveprefix = {arXiv},
  copyright     = {arXiv.org perpetual, non-exclusive license},
  eprint        = {2511.05208},
  primaryclass  = {cond-mat.mes-hall},
  publisher     = {American Physical Society (APS)},
}

@misc{Maiani_2026_Code,
  author       = {Maiani, Andrea and Wozny, Simon and Abadillo-Uriel, Jos{\'e} Carlos and Cortijo, Alberto and Aguado, Ram{\'o}n and Seoane Souto, Rub{\'e}n},
  title        = {{Spin-selective tunneling in altermagnetic quantum dots: code and data}},
  year         = {2026},
  howpublished = {Zenodo},
  url          = {https://doi.org/10.5281/zenodo.0000000},
  publisher    = {Zenodo},
  doi          = {10.5281/zenodo.0000000},
}

@misc{Mao_2026_Spin,
  author        = {Mao, Yue and Zhuang, Yu-Chen and Miao, Cheng-Ming and Sun, Yu-Fei and Sun, Qing-Feng},
  title         = {Spin-textured orbitals in altermagnetic artificial atoms},
  year          = {2026},
  month         = {aug},
  url           = {https://arxiv.org/abs/2608.29861},
  eprint        = {2608.29861},
  archiveprefix = {arXiv},
  primaryclass  = {cond-mat.mes-hall},
}

@article{Mencos_Nature2026,
  author    = {Mencos, Jone and Badura, Antonin and Dolan, Eoin and Beckert, Sebastian and Gonz{\'a}lez-Hern{\'a}ndez, Rafael and Sigales, Nicol{\'a}s and Kokkeler, Tim and Kounta, Isma{\"\i}la and Petit, Matthieu and Guillemard, Charles and Hellenes, Anna Birk and Campos, Warlley and Rial, Javier and Kriegner, Dominik and Baltz, Vincent and Hueso, Luis E. and Sinova, Jairo and Bergeret, F. Sebastian and Gomonay, Olena and Jungwirth, Tom{\'a}{\v s} and {\v S}mejkal, Libor and Michez, Lisa and Reichlova, Helena and Casanova, F{\`e}lix},
  title     = "Time-reversal-symmetry-breaking spin injection from a
               compensated magnet",
  journal   = "Nature",
  year      = 2026,
  month     = oct,
  doi       = {10.1038/s41586-026-11227-7},
  url       = {https://www.nature.com/articles/s41586-026-11227-7},
  publisher = "Springer Science and Business Media LLC",
  copyright = "https://www.springernature.com/gp/researchers/text-and-data-mining",
  language  = "en",
}

@article{Sigales_PRB2026,
  author    = {Sigales, Nicolas and Kokkeler, Tim and De Polsi, Gonzalo and Bergeret, F. Sebastian},
  title     = {Spin-splitter and inverse effects in altermagnetic hybrid structures},
  journal   = {Phys. Rev. B},
  year      = {2026},
  volume    = {113},
  number    = {18},
  pages     = {184447},
  numpages  = {11},
  month     = {May},
  doi       = {10.1103/72xm-5w1y},
  url       = {https://link.aps.org/doi/10.1103/72xm-5w1y},
  publisher = {American Physical Society},
}

@misc{Tjernshaugen_2026_Superconducting,
  author        = {Tjernshaugen, Johanne Bratland and Amundsen, Morten and Linder, Jacob},
  title         = {Superconducting Qubits with Altermagnetic Josephson Junctions},
  year          = {2026},
  url           = {https://arxiv.org/abs/2606.02761},
  doi           = {10.48550/arXiv.2606.02761},
  publisher     = {arXiv},
  copyright     = {arXiv.org perpetual, non-exclusive license},
  eprint        = {2606.02761},
  archiveprefix = {arXiv},
  primaryclass  = {quant-ph},
}

@misc{Vakili_2026_Gate,
  author        = {Vakili, Hamed},
  title         = {Gate-Controlled Spin Qubits in Confined Altermagnets},
  year          = {2026},
  url           = {https://arxiv.org/abs/2606.24150},
  archiveprefix = {arXiv},
  eprint        = {2606.24150},
  primaryclass  = {cond-mat.mes-hall},
}

@misc{Yang_2026_Visualizing,
  author        = {Yang, Guofei and Li, Chuang and Wang, Chengwei and Zhao, Xudong and Wan, Yifan and Gui, Hengrui and Zeng, Guoqing and Cao, Saizheng and Hu, Chuqiao and Yu, Qihe and Zhang, Yujia and Chen, Dong and Liu, Yu and Song, Yu and Zhang, Yongjun and Liu, Fei and Hu, Lun-Hui and Jiao, Lin and Yuan, Huiqiu},
  title         = {Visualizing spin-polarization of an altermagnet {KV$_2$Se$_2$O} via spin-selective tunneling},
  year          = {2026},
  month         = {mar},
  url           = {https://arxiv.org/abs/2603.21969},
  eprint        = {2603.21969},
  archiveprefix = {arXiv},
  primaryclass  = {cond-mat.mtrl-sci},
}

@misc{SM,
  note = {See Supplemental Material at [URL will be inserted by publisher] for the continuum dot and contact model, the hybridization matrix with spin--orbit coupling, the nonequilibrium Green's function and interacting transport formalism, and the orthogonal-contact geometry}
}

\end{document}


\title{Supplemental Material: Spin-selective tunneling in altermagnetic quantum dots}

\maketitle

\tableofcontents

\section{Continuum dot and contact model}
\label{sec:continuum-model-and-static-dot-basis}

We consider a dot electrostatically defined through an anisotropic harmonic confinement
in a two-dimensional electron gas with a $d$-wave altermagnetic term. This can be modeled
with a minimal Hamiltonian
\begin{equation}
  \label{eq:continuum-hamiltonian}
  H = \varepsilon_g
      + \frac{p_x^2 + p_y^2}{2m}
      + \frac{m}{2}\left( \omega_x^2 x^2 + \omega_y^2 y^2 \right)
      + \beta_d\left( p_x^2 - p_y^2 \right)\sigma_3
      + \frac{h_z}{2}\,\sigma_3
      + \alpha\left( \sigma_2 p_x - \sigma_1 p_y \right),
\end{equation}
with plunger-gate energy $\varepsilon_g$, effective mass $m$, confinement
frequencies $\omega_\nu = 1/(m\ell_\nu^2)$, with $\omega_x = \omega_y \equiv \omega_0$ and
$\ell_x = \ell_y \equiv \ell$ for a circular dot, altermagnetic coefficient $\beta_d$, longitudinal
splitting $h_z$, and Rashba coupling $\alpha$. We set $\hbar = 1$ and define the nondimensional
$b \equiv m\beta_d$.

The two-dimensional altermagnetic term describes the lowest confined subband of a film. It
requires the spin splitting, projected onto that subband, to be even, in-plane, and
sign-changing. This holds when the operations exchanging the two sublattices act on the in-plane
momentum.

We write $H = \varepsilon_g + H_0 + V_R$, where $H_0$ contains the kinetic, confinement,
altermagnetic, and longitudinal-field terms, and $V_R = \alpha(\sigma_2p_x - \sigma_1p_y)$
is the relativistic spin--orbit coupling. Since $[H_0,\sigma_3] = 0$, we solve the
two spin sectors first and then represent $V_R$ in that basis.

\subsection{Finite-\texorpdfstring{$\beta_d$}{beta\_d} eigenbasis of the isolated dot}
\label{sec:finite-beta-eigenbasis-of-the-isolated-dot}

For a fixed spin projection $s = \pm1$ the dot without $V_R$ is an anisotropic oscillator with
spin-dependent kinetic coefficients $(1 \pm 2sb)/2m$ along $x$ and $y$. Its normal-mode frequencies
and oscillator lengths are
\begin{equation}
  \label{eq:spin-sector-lengths-finite-beta}
  \Omega_{x,s} = \omega_x\sqrt{1 + 2sb}, \quad \Omega_{y,s}
               = \omega_y\sqrt{1 - 2sb}, \qquad \ell_{x,s}
               = \ell_x\left( 1 + 2sb \right)^{1 / 4}, \quad \ell_{y,s}
               = \ell_y\left( 1 - 2sb \right)^{1 / 4},
\end{equation}
and the product states $\ket{n_x\,n_y\,s}$ have energies
$E_{n_x n_y s} = \Omega_{x,s}(n_x + \tfrac12) + \Omega_{y,s}(n_y + \tfrac12) + sh_z/2$. Each is a
spinor $\ket{s}$ times an orbital that depends on $s$ through the squeezed lengths,
\begin{equation}
  \label{eq:finite-beta-orbital-wavefunctions}
  \phi^{(s)}_{n_x n_y}(x,y) = \frac{H_{n_x}\!\left( x/\ell_{x,s} \right) H_{n_y}\!\left( y/\ell_{y,s} \right)}{\sqrt{2^{n_x+n_y}n_x!n_y!\,\pi\,\ell_{x,s}\ell_{y,s}}} \exp\!\left[ -\frac{x^2}{2\ell_{x,s}^2} - \frac{y^2}{2\ell_{y,s}^2} \right],
\end{equation}
so the ground state in each spin sector is a squeezed Gaussian.

For circular confinement the ground doublet is exactly unsplit at $h_z = \alpha = 0$, while states
with $n_x \ne n_y$ have a spin splitting
$\Delta_{n_x,n_y} = E_{n_x n_y+} - E_{n_x n_y-} = h_z + 2\beta_d(n_x - n_y)/\ell^2 + O(\beta_d^3)$.
For $|b| \ll 1$, $\ell_{x,s} = \ell_x(1 + sb/2)$ and $\ell_{y,s} = \ell_y(1 - sb/2)$ up to $O(b^2)$.

\begin{figure}
  \centering
  \includegraphics[width=\linewidth]{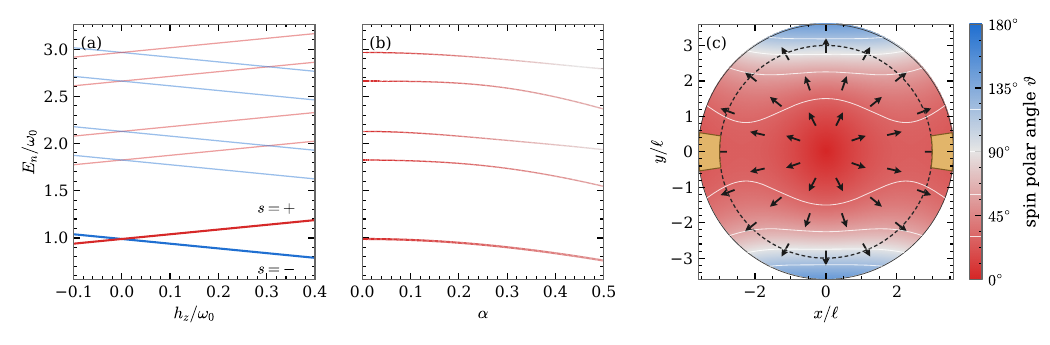}
  \caption{
    The ground doublet and the spin texture a contact samples. (a) Spectrum against the out-of-plane
    field at $\alpha = 0$, where $\sigma_3$ is conserved and every level has a definite N\'eel
    spin (red $s = +$, blue $s = -$; thick lines: ground doublet). The field splits the doublet
    by $h_z$ itself; the first excited shell is already split at $h_z = 0$ by the altermagnetic
    term. (b) Spectrum against the Rashba coupling at $h_z = 0$. Every level stays exactly doubly
    degenerate; each pair is colored by the polar angle
    $\vartheta = \arccos\langle\sigma_3\rangle$ of the spin of its $s = +$ member, which for the
    doublet grows linearly in $\alpha$ ($1 - \langle\sigma_3\rangle \propto \alpha^2$): Rashba
    coupling mixes the spins but cannot split the doublet. (c) The $s = +$ state of the
    lowest-energy doublet for $\eta = 0.15$: polar angle $\vartheta = \arccos(S_z/n)$ of the local
    spin (color, same scale as b; white contours $30^\circ$), direction of the in-plane spin
    $\bm S_\perp$ (arrows), the contact circle $r = R_{\mathrm{lead}}$ (dashed), and a
    same-axis pair of leads. The spin tilts away from the N\'eel axis toward the
    radial direction as the distance grows, reaching $\vartheta = 25^\circ$ at the
    contact on the $x$ axis, of order $\eta\rho$. Along $y$, where the $s = +$
    orbital is the narrower one, the admixed $s = -$ tail dominates beyond the
    contact radius and the local spin reverses. Parameters: $m = \omega_0 = 1$
    (circular dot, so $\ell = 1$ and $\eta = \alpha$), $b = 0.15$, $\rho = 3$, $\varphi = 0.15$.
  }
  \label{fig:spectrum}
\end{figure}

We represent the full single-particle Hamiltonian in the exact finite-$\beta_d$ basis
$\ket{n_x\,n_y\,s}$, truncated at $0 \le n_x,n_y \le N_{\mathrm{cut}}$, and diagonalize it
numerically. Since $V_R$ flips the $\sigma_3$ projection,
\begin{equation}
  \label{eq:same-spin-matrix-elements-vanish}
  \bra{r_x\,r_y\,s} V_R \ket{n_x\,n_y\,s} = 0,
\end{equation}
it couples only opposite spin sectors, through overlap and momentum matrix elements between
oscillator states of different widths, which are known in closed form. Figure~\ref{fig:spectrum}
shows the resulting spectrum.

\subsection{Hybridization matrix for a wedge-shaped contact}
\label{sec:derivation-of-the-hybridization-matrix}

\begin{figure}
  \centering
  \includegraphics[width=\linewidth]{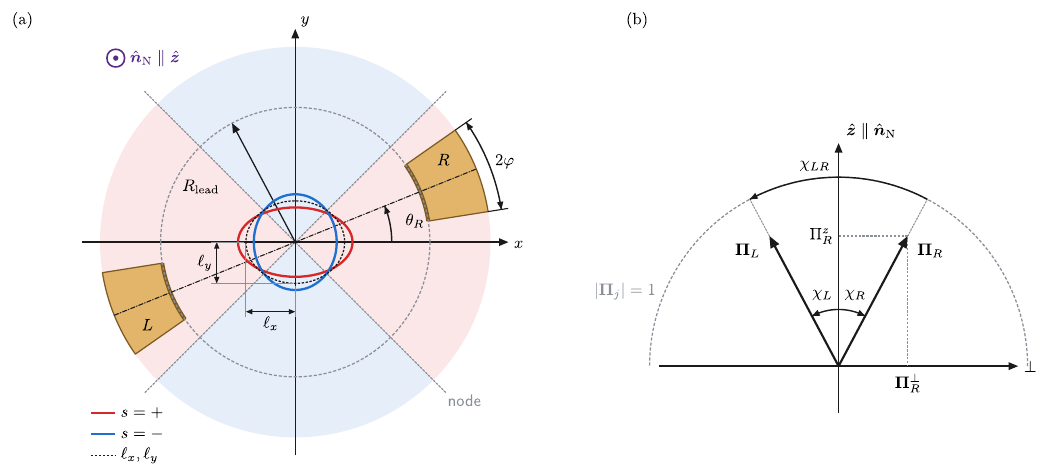}
  \caption{
    Contact geometry. (a) Top view in the crystal frame. The tint marks the sign of the $d$-wave
    form factor $\cos2\theta$ (red positive, blue negative), with nodes on the diagonals. The
    inner dashed circle represents the QD confinement, characterized by lengths $\ell_x$
    and $\ell_y$; the red and blue contours are the $s = \pm$ ground-state
    densities, of lengths $\ell_{x,s}$ and $\ell_{y,s}$. Each lead occupies the
    wedge $r > R_{\mathrm{lead}}$, $|\theta - \theta_j| < \varphi$, and couples
    through the arc at $r = R_{\mathrm{lead}}$ (outer dashed line). The leads form a same-axis
    pair, $\theta_L = \theta_R + \pi$, and the N\'eel vector $\hat{\bm n}$ points
    out of the plane. (b) Contact polarizations $\bm\Pi_j$ in pseudospin space, tilted by
    $\chi_j$ from the N\'eel axis by Rashba coupling, with relative angle $\chi_{LR}$.
    Proportions are exaggerated for clarity.
  }
  \label{fig:contact-geometry}
\end{figure}

We choose a wedge-shaped lead; its specific shape does not affect the presented results
for a sufficiently small angular opening. At $\alpha = 0$, spin is conserved and the dot
eigenstates are the squeezed product states
$\psi_{n_x n_y,s} = \phi^{(s)}_{n_x n_y}$ of
Eq.~\eqref{eq:finite-beta-orbital-wavefunctions}, with lengths given by
Eq.~\eqref{eq:spin-sector-lengths-finite-beta}. Lead $j$ occupies the exterior wedge
$r > R_{\mathrm{lead}}$, $|\theta - \theta_j| < \varphi$, where $\theta_j$ is measured from
the crystal $x$ axis and $\varphi$ is the half opening angle
[Fig.~\ref{fig:contact-geometry}(a)]. For local spin-conserving tunneling to a wide-band lead
through the arc at $r = R_{\mathrm{lead}}$, the hybridization matrix is
\begin{equation}
  \label{eq:exact-hybridization-arc}
  \Gamma^{(\theta_j,\varphi)}_{n_x n_y s,\;n_x' n_y' s'} = \Gamma_{\mathrm t}\,\delta_{ss'}\,R_{\mathrm{lead}} \int_{\theta_j-\varphi}^{\theta_j+\varphi} \dd\theta\, \psi^\ast_{n_x n_y,s}(R_{\mathrm{lead}},\theta)\, \psi_{n_x' n_y',s}(R_{\mathrm{lead}},\theta),
\end{equation}
where
$\psi_{n_x n_y,s}(R,\theta) = \phi_{n_x}^{(x,s)}(R\cos\theta)\,\phi_{n_y}^{(y,s)}(R\sin\theta)$
and $\Gamma_{\mathrm t} = 2\pi |t|^2\mathcal N(E_F)$, with tunneling amplitude $t$ and lead density
of states $\mathcal N(E_F)$ at the Fermi level.

For the lowest-energy doublet,
$\Lambda_s = R_{\mathrm{lead}}^2(\ell_{x,s}^{-2} + \ell_{y,s}^{-2})/2$ and
$\delta_s = R_{\mathrm{lead}}^2(\ell_{x,s}^{-2} - \ell_{y,s}^{-2})/2$. Then
\begin{equation}
  \label{eq:ground-state-hybridization-exact}
  \Gamma^{(\theta_j,\varphi)}_{00,s} = \frac{\Gamma_{\mathrm t} R_{\mathrm{lead}}}{\pi \ell_{x,s}\ell_{y,s}} e^{-\Lambda_s} \mathcal{I}_s(\theta_j,\varphi),
\end{equation}
where
\begin{equation}
  \label{eq:angular-integral-ground-state}
  \mathcal{I}_s(\theta_j,\varphi) \equiv \int_{\theta_j-\varphi}^{\theta_j+\varphi} \dd\theta\, e^{-\delta_s \cos 2\theta}.
\end{equation}
For a narrow symmetric wedge, evaluating the integrand at the contact center gives, up to
$O(\varphi^3)$,
\begin{equation}
  \label{eq:ground-state-hybridization-narrow-wedge}
  \Gamma^{(\theta_j,\varphi)}_{00,s} \approx \frac{2\Gamma_{\mathrm t} R_{\mathrm{lead}}\varphi}{\pi \ell_{x,s}\ell_{y,s}} \exp\!\left[ -\Lambda_s - \delta_s\cos 2\theta_j \right].
\end{equation}

For circular confinement, $\ell_x = \ell_y \equiv \ell$, weak altermagnetism, $|b| \ll 1$,
and $\rho \equiv R_{\mathrm{lead}}/\ell$, one has $\Lambda_s = \rho^2 + O(b^2)$ and
$\delta_s = -sb\rho^2 + O(b^3)$, so in the narrow-arc limit
\begin{equation}
  \label{eq:ground-state-hybridization-weak-beta-narrow}
  \Gamma^{(\theta_j,\varphi)}_{00,s}
  \approx \Gamma_0 \exp\!\left[ s\,b\rho^2\cos 2\theta_j \right], \qquad \Gamma_0
  \equiv  \frac{2\Gamma_{\mathrm t} R_{\mathrm{lead}}\varphi}{\pi \ell^2} e^{-\rho^2}.
\end{equation}

\subsection{Contact asymmetry beyond parabolic confinement}
\label{sec:beyond-parabolic}

The exponent $A = b\rho^2\cos2\theta_j$ of
Eq.~\eqref{eq:ground-state-hybridization-weak-beta-narrow} comes from the Gaussian tail of the
harmonic orbital, originated from the leads sampling of the exponential tail of the
QDs' wavefunctions. The structure of $A$ does not depend on the specific form of the
confining potential.For instance, taking $\alpha = 0$ and a rotationally symmetric
confinement $V(r)$, so that spin sector $s$ has
$H_s = (1 + 2sb)k_x^2/2m + (1 - 2sb)k_y^2/2m + V(r)$, and write its nodeless ground state as
$\psi_s = e^{-W_s}$. The Schr\"odinger equation becomes the Riccati equation
\begin{equation}
  \label{eq:riccati-ground-state}
  \frac{1}{2m}\sum_{i=x,y}\left( 1 \pm 2sb \right)\left[ (\partial_i W_s)^2 - \partial_i^2 W_s \right]
  = V(r) - E_s,
\end{equation}
with the upper sign for $x$. The $C_4\mathcal T$ symmetry gives $E_+ = E_-$, and
$E_s = E_0 + O(b^2)$. To first order, $W_s = W_0(r) - s\,b\,g(r)\cos2\theta$, since the source term
generated by $b$, $(\partial_x W_0)^2 - (\partial_y W_0)^2 - (\partial_x^2 - \partial_y^2)W_0$,
is a pure $\cos2\theta$ harmonic. The radial function obeys
\begin{equation}
  \label{eq:riccati-radial}
  W_0' g' - \frac12\left( g'' + \frac{g'}{r} - \frac{4g}{r^2} \right)
  = W_0'^2 - \left( W_0'' - \frac{W_0'}{r} \right).
\end{equation}
For the harmonic dot, $W_0 = r^2/2\ell^2$, the solution is $g = W_0$ exactly. In the tail, where the
decay constant $\kappa$ satisfies $\kappa r \gg 1$ and varies slowly, the derivative corrections
are subleading and $g' \simeq W_0'$. Interior matching fixes an additive constant in $g$, so the
following relation holds at leading tunneling action, up to an $O(b)$ offset. The contact asymmetry
$A = \frac12\ln(|\psi_+|^2/|\psi_-|^2)$ at the contact is therefore
\begin{equation}
  \label{eq:contact-asymmetry-wkb}
  A(R_{\mathrm{lead}},\theta_j) \simeq b\cos2\theta_j\,\ln\frac{|\psi_0(0)|^2}{|\psi_0(R_{\mathrm{lead}})|^2}
                                \equiv b\cos2\theta_j\,\ln\frac{1}{\mathcal T},
\end{equation}
that is, $b\cos2\theta_j$ times twice the leading bare action $S_0$ in the classically forbidden
tail. Equivalently, $\Gamma_\pm \propto \mathcal T^{\,1 \mp b\cos2\theta_j}$: the two spins see
the bare tail suppression raised to a spin- and angle-dependent power. The harmonic dot gives
$\ln(1/\mathcal T) = \rho^2$ and recovers
Eq.~\eqref{eq:ground-state-hybridization-weak-beta-narrow}. A flat-bottom well of radius $r_d$
behind a barrier of height $V_0 - E$ gives
$A \simeq 2b\kappa_0\,(R_{\mathrm{lead}} - R^\ast)\cos2\theta_j$, with
$\kappa_0 = \sqrt{2m(V_0 - E)}$ and an effective origin $R^\ast$ inside the well.

At finite contact width the rates must be integrated over the arc before taking their logarithmic
ratio. Prefactors of the tunneling matrix element add a further $O(b)$ offset that does not grow
with the distance, and if the barrier is not itself altermagnetic, only the action accumulated
inside the altermagnet enters $A$. An elliptic dot splits the doublet, $E_+ \neq E_-$,
and adds an isotropic contribution
$\delta A \approx m(E_+ - E_-)(R_{\mathrm{lead}} - R^\ast)/\kappa_0$, the lower spin
decaying faster; this is the level-splitting channel of the main text, separate from the form
factor. With Rashba coupling, the spin sectors mix, and the dressing of
Sec.~\ref{sec:soc-doublet-block} applies on top of $A$.

\section{Hybridization matrix in the presence of spin--orbit coupling}
\label{sec:soc-hybridization}

We now reinstate the Rashba term $V_R = \alpha\left( \sigma_2 p_x - \sigma_1 p_y \right)$ and
show that, in the regime where altermagnetism dominates over the relativistic spin--orbit
coupling, its entire effect on transport at first order in $\alpha$ is a dressing of the
hybridization matrix: the dot spectrum and the N\'eel-direction basis of
Sec.~\ref{sec:finite-beta-eigenbasis-of-the-isolated-dot} are unchanged, and the
couplings acquire a spin-flip component. Throughout this section the altermagnetic
coupling $\beta_d$ is kept exactly through the squeezed basis, while the small
parameters are $m\alpha\ell$ and $m\alpha R_{\mathrm{lead}}$.

\subsection{Dressed coupling and contact spin-flip amplitude}
\label{sec:soc-dressed-coupling}

In the N\'eel basis $\ket{n_x\,n_y\,s}$ the Rashba term is spin-off-diagonal,
Eq.~\eqref{eq:same-spin-matrix-elements-vanish}, and orbital parity makes its matrix elements
within the isolated ground doublet vanish. Away from crossings with excited states, the energies
therefore change only at $O(\alpha^2)$, while the first-order state correction is encoded in the
anti-Hermitian admixture matrix
\begin{equation}
  \label{eq:soc-admixture-matrix}
  S_{nn'} = \frac{\alpha\bra{n}H_{\alpha}\ket{n'}}{E_{n'} - E_n}, \qquad n \neq n',
\end{equation}
where $H_\alpha = V_R/\alpha$ and the retained doublet is excluded from the admixture.
Near an excited-state degeneracy, the degenerate block must instead be diagonalized.
Since transport observables are basis-invariant traces, the state correction can be absorbed into
\begin{equation}
  \label{eq:soc-dressed-gamma}
  \Gamma_{\mathrm{eff}} = \left( 1 + S \right)^\dagger \Gamma^{(\theta_j,\varphi)} \left( 1 + S \right)
                        = \Gamma^{(\theta_j,\varphi)}
                          + \left[ \Gamma^{(\theta_j,\varphi)},S \right]
                          + O(\alpha^2),
\end{equation}
while the Hamiltonian entering the Green's functions of Sec.~\ref{sec:negf-formalism}
keeps its $\alpha = 0$ eigenvalues. Since $\Gamma^{(\theta_j,\varphi)}$ is spin-diagonal and $S$
is spin-off-diagonal, the commutator is purely spin-flip. Thus the diagonal rates of
Eq.~\eqref{eq:ground-state-hybridization-weak-beta-narrow} and the main-text transmission,
conductance, and peak-height asymmetry change only at $O(\alpha^2)$; the first-order correction
produces a contact spin texture, as in altermagnetic artificial
atoms~\cite{Mao_2026_Spin}.

\subsection{Ground-doublet coupling block and effective polarization}
\label{sec:soc-doublet-block}

Within the ground doublet, the dressed coupling matrix of a narrow contact at $h_z = 0$ takes
the form
\begin{equation}
  \label{eq:soc-doublet-block}
  \Gamma^{(00)} = \Gamma_0 \begin{pmatrix}
                             e^{b\rho^2\cos2\theta_j} & \zeta(\theta_j)           \\
                             \zeta^\ast(\theta_j)     & e^{-b\rho^2\cos2\theta_j}
                           \end{pmatrix},
\end{equation}
where $\Gamma_\pm \equiv \Gamma_0e^{\pm b\rho^2\cos2\theta_j}$ denote the diagonal
entries and $\Gamma_0\zeta$ is the spin-flip matrix element. At $b = 0$, the two
first-order contributions to this element cancel exactly. The spin-flip coupling
survives only through the mismatch of the differently squeezed spin clouds.
Evaluating Eq.~\eqref{eq:soc-dressed-gamma} with the admixture matrix of
Eq.~\eqref{eq:soc-admixture-matrix} gives, to leading order in $b$,
\begin{equation}
  \label{eq:soc-off-diagonal-element}
  \zeta(\theta_j)
  =      \frac{4}{3}\eta b\rho \left[ e^{i\theta_j} - \rho^2\cos2\theta_j\,e^{-i\theta_j} \right], \qquad \eta
  \equiv m\alpha\ell.
\end{equation}
The spin-flip element tilts the contact polarization from the N\'eel axis by
$\chi_j = \operatorname{atan2}(|\zeta_j|, \sinh A_j)$, which to leading order in $b$ is
\begin{equation}
  \label{eq:soc-mixing-angle}
  \tan\chi_j
  = \frac{2\Gamma_0|\zeta_j|}{\Gamma_+ - \Gamma_-}
  = \frac{4\eta}{3\rho\cos2\theta_j} \left| e^{i\theta_j} - \rho^2\cos2\theta_j\,e^{-i\theta_j} \right|.
\end{equation}
The last form linearizes in $b$ at fixed $\rho$ and is not uniform for large $b\rho^2$. Both
$\Gamma_0\zeta_j$ and $\Gamma_+ - \Gamma_-$ are linear in $b$, so the spin tilt is of order $\eta$
and \emph{independent of} $b$. It becomes complete, $\chi_j \to \pi/2$, at the $d$-wave nodes
$\theta_j \to \pi / 4$, where $\Gamma_+ = \Gamma_-$: spin-selective coupling is most fragile to
spin--orbit coupling precisely where the altermagnetic anisotropy vanishes.

The projected hybridization of contact $j = L,R$ is a positive Hermitian $2 \times 2$ matrix.
With the altermagnetic exponent $A_j \equiv b\rho^2\cos2\theta_j$, the Pauli decomposition of
Eq.~\eqref{eq:soc-doublet-block} reads
\begin{equation}
  \label{eq:sm-effective-polarization}
  \Gamma_j
  = \bar{\Gamma}_j \left( \bm{1} + \bm{\Pi}_j \cdot \bm{\sigma} \right), \qquad \bar{\Gamma}_j
  = \frac{1}{2}\tr\Gamma_j
  = \Gamma_{0j}\cosh A_j, \qquad \bm{\Pi}_j
  = \frac{\left( \operatorname{Re}\zeta_j, -\operatorname{Im}\zeta_j, \sinh A_j \right)}{\cosh A_j}.
\end{equation}
The altermagnetic exponent enters only the component along the N\'eel axis $\hat{\bm z}$,
$\Pi^z_j = \tanh A_j$, the only one left at $\eta = 0$; the spin-flip amplitude $\zeta_j$ of
Eq.~\eqref{eq:soc-off-diagonal-element} supplies the transverse components, and we write
$\Pi_j = |\bm{\Pi}_j|$. Numerical results use the full projected matrices, which hold only for
small $\eta\rho$. Since $\zeta_j$ carries the contact-dependent phase
$e^{i\theta_j}$, the tilt differs from contact to contact, and two leads at
different angles acquire non-collinear polarization axes. The tilt of $\bm{\Pi}_j$ from the N\'eel
axis is the angle $\chi_j$ of
Eq.~\eqref{eq:soc-mixing-angle} [Fig.~\ref{fig:contact-geometry}(b)]. We use $\theta$ for angles
measured from the crystal axes in real space and $\chi$ for angles in pseudospin space. Note that
$\bm{\Pi}_j$ is defined only up to the arbitrary basis choice inside the degenerate doublet, which
rotates every $\bm{\Pi}_j$ simultaneously; only $\Pi_j$ and the relative angle
\begin{equation}
  \label{eq:sm-relative-angle}
  \chi_{LR} = \arccos\frac{\bm{\Pi}_L \cdot \bm{\Pi}_R}{\left| \bm{\Pi}_L \right|\left| \bm{\Pi}_R \right|}
\end{equation}
are basis independent. For a same-axis pair the $C_{2z}$ rotation maps one contact onto the other
and reverses the transverse components of $\bm\Pi$, so the two tilts are opposite.
For $A_j > 0$, $\chi_{LR} = 2\chi_j$.

Time-reversal symmetry constrains the contact polarization more strongly. For a Kramers
pair choose $\mathcal T = i\sigma_y\mathcal K$. If tunneling to each normal reservoir
preserves time reversal, its Hermitian projected hybridization satisfies
$\Gamma_j = \sigma_y\Gamma_j^*\sigma_y$. Substitution of
$\Gamma_j = \bar\Gamma_j(\bm1 + \bm\Pi_j \cdot \bm\sigma)$ then gives $\bm\Pi_j = 0$.

\section{Noninteracting transport and current polarization}
\label{sec:negf-formalism}

\subsection{Charge and spin currents}

For noninteracting transport we use the standard nonequilibrium Green's function expressions in
the wide-band limit. The retarded Green's function in the truncated dot basis is
\begin{equation}
  \label{eq:retarded-greens-function}
  G^R(E) = \left[ (E + i0^+)\bm{1} - H_{\mathrm{dot}} - \Sigma^R(E) \right]^{-1}, \qquad \Sigma^R
         = -\frac{i}{2}\sum_j\Gamma^{(j)},
\end{equation}
with $H_{\mathrm{dot}}$ the truncated Hamiltonian of
Sec.~\ref{sec:finite-beta-eigenbasis-of-the-isolated-dot}. The transmission between leads
$j$ and $k$ is given by the Caroli formula,
\begin{equation}
  \label{eq:caroli-formula}
  T_{jk}(E) = \tr\!\left[ \Gamma^{(j)} G^R(E) \Gamma^{(k)} G^A(E) \right],
\end{equation}
and the current follows from the Landauer--B\"uttiker formula,
\begin{equation}
  \label{eq:landauer-buettiker-current}
  I_j = \frac{e}{h} \sum_k \int \dd E\, T_{jk}(E) \left[ f_j(E) - f_k(E) \right].
\end{equation}

We resolve the current into the drain by the spin of the outgoing electron. Writing
the arc hybridization of Eq.~\eqref{eq:exact-hybridization-arc}, extended to spinor
wavefunctions, as $\Gamma^{(j)} = C^{(j)}C^{(j) \dagger}$, with
$C^{(j)}_{i,(\theta\sigma)} = \sqrt{\Gamma_{\mathrm t}R_{\mathrm{lead}}}\,\psi^\ast_{i,\sigma}(R_{\mathrm{lead}},\theta)$
the amplitude for dot state $i$ to tunnel into the lead channel at angle $\theta$ with spin
$\sigma$ along the N\'eel axis, the spin current into the drain is
\begin{equation}
  \label{eq:sm-spin-current}
  I_s^\mu = \frac{e}{h}\int\dd E\,\tr\!\left[ \Gamma^{(R)}_\mu G^R(E)\Gamma^{(L)}G^A(E) \right] \left[ f_L(E) - f_R(E) \right], \qquad \Gamma^{(R)}_\mu
          = C^{(R)}\left( \bm 1_\theta \otimes \sigma_\mu \right)C^{(R) \dagger},
\end{equation}
which reduces to the charge current $I_c$ of Eq.~\eqref{eq:landauer-buettiker-current} when
$\sigma_\mu$ is replaced by the identity. The current polarization is $\bm P = \bm I_s / I_c$.

At $\alpha = 0$, $\Gamma^{(R)}_z$ is spin diagonal and $\Gamma^{(R)}_{x,y}$ are spin off-diagonal
in the N\'eel basis. At first order in $\alpha$, the dressing of Eq.~\eqref{eq:soc-dressed-gamma} is
spin off-diagonal, and so is the change of $G^R$ through the self-energy, while the dot
energies change only at $O(\alpha^2)$. In the trace of
Eq.~\eqref{eq:sm-spin-current}, a single spin-off-diagonal factor between spin-diagonal ones
gives zero. Hence $I_c$ and $I_s^z$ change only at $O(\eta^2)$, whereas the transverse components,
whose weights $\Gamma^{(R)}_{x,y}$ are themselves spin off-diagonal, start at $O(\eta)$.

\subsection{Resonance weights and spin selectivity}
\label{sec:current-polarization}

For an isolated spin-conserving resonance, define its weight
$W_s = \Gamma_{L,s}\Gamma_{R,s}/\Gamma_s$, with $\Gamma_s = \Gamma_{L,s} + \Gamma_{R,s}$.
Integration over the full line gives
\begin{equation}
  \label{eq:sm-step-current}
  I_s = \frac{e}{h}\int\dd E\,T^{(s)}_{LR}(E) = \frac{e}{\hbar}\,W_s,
\end{equation}
which leads to the current plateau shown in Fig.~3 of the main text.

For spin-conserving contacts, the zero-temperature peak transmission is spin
independent (unity for identical contacts), so the asymmetry is a thermal effect. For
$k_{\mathrm B}T \gg \Gamma_s$ the conductance samples the integrated weight of the Lorentzian,
\begin{equation}
  \label{eq:sm-peak-weight}
  G^{\mathrm{max}}_s
  \sim \frac{e^2}{h}\, \frac{W_s}{k_{\mathrm B}T}, \qquad W_s
  =    \frac{\Gamma_{0L}\Gamma_{0R}\, e^{s\left( A_L+A_R \right)}}{\Gamma_{0L} e^{sA_L} + \Gamma_{0R} e^{sA_R}},
\end{equation}
with $\Gamma_{j,s} = \Gamma_{0j} e^{sA_j}$ from
Eq.~\eqref{eq:sm-effective-polarization}.

At $\alpha = 0$, for a \emph{same-axis} pair with $A_L = A_R \equiv A$, the thermal peak weight in
Eq.~\eqref{eq:sm-peak-weight} factorizes into $\Gamma_{0L}\Gamma_{0R}/(\Gamma_{0L} + \Gamma_{0R})$
times $e^{sA}$, and the prefactor cancels from the peak-height asymmetry,
\begin{equation}
  \label{eq:sm-peak-asymmetry}
  \mathcal{A} \equiv \frac{G^{\mathrm{max}}_{+} - G^{\mathrm{max}}_{-}}{G^{\mathrm{max}}_{+} + G^{\mathrm{max}}_{-}}
              =      \tanh A
              =      \Pi^z,
\end{equation}
for any ratio $\Gamma_{0R}/\Gamma_{0L}$. Differences in barrier action or contact position,
$A_L \neq A_R$, do not cancel. An orthogonal pair (Sec.~\ref{sec:orthogonal-contacts}) has
$A_R = -A_L$. With $\Gamma_{0L} = \Gamma_{0R}$, the weight $\Gamma_0/2\cosh A$ is independent of
$s$, so the asymmetry vanishes, and the total peak weight falls below the same-axis value by
\begin{equation}
  \label{eq:sm-thermal-valve-ratio}
  \frac{G^{\mathrm{max}}_{\mathrm{AP}}}{G^{\mathrm{max}}_{\mathrm{P}}} = \frac{1}{\cosh^2 A}
                                                                       = 1 - \Pi^2,
\end{equation}
where AP (P) labels the orthogonal (same-axis) pair, whose polarizations $\Pi^z_j$ are
antiparallel (parallel).

\begin{figure}[t]
  \centering
  \includegraphics[width=\linewidth]{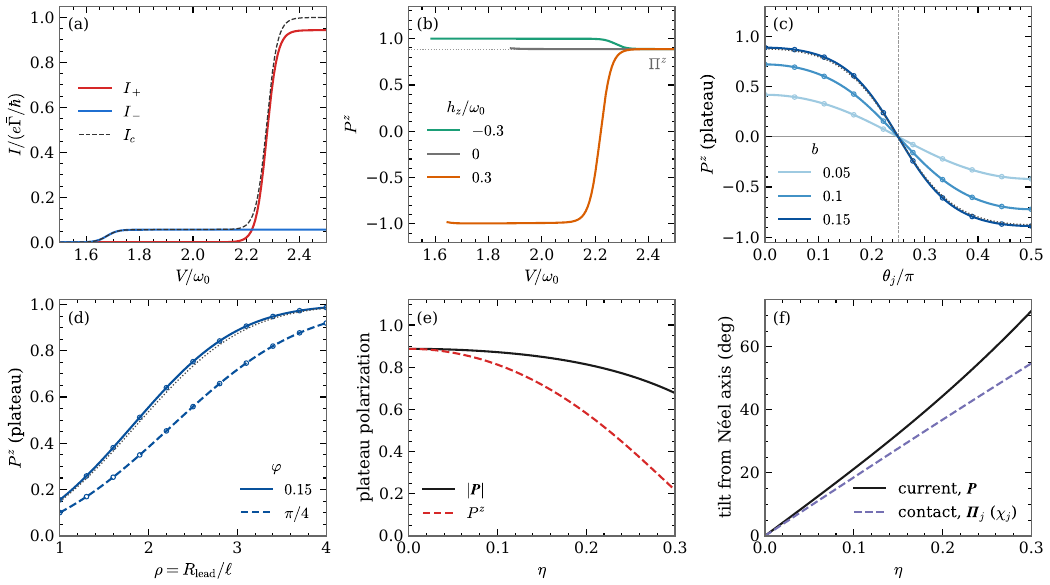}
  \caption{
    Spin polarization of the current through the ground doublet, same-axis contacts. (a)
    N\'eel-spin-resolved currents and charge current against bias for $h_z = 0.3\,\omega_0$: the
    lower level carries the weakly coupled spin, the second step the strongly coupled one. (b)
    Current polarization $P^z$ against bias for three Zeeman splittings, drawn where the current
    exceeds $1\%$ of its plateau value: $\pm 1$ between the steps of a resolved doublet and $\Pi^z$
    (dotted) once both levels conduct. (c) Plateau polarization against the axis
    $\theta_j$ of the pair for three $b$: closed form
    Eq.~\eqref{eq:sm-step-polarization} (lines), microscopic doublet (circles), and
    $\tanh(b\rho^2\cos2\theta_j)$ (dotted, $b = 0.15$); the dashed line marks the node. (d) Plateau
    polarization against $\rho$ for a narrow and a wide contact, with $\tanh(b\rho^2)$ dotted. (e)
    Magnitude and N\'eel component of the plateau polarization against the Rashba strength, and (f)
    the tilt of $\bm P$ from the N\'eel axis, next to the tilt $\chi_j$ of the contact
    polarization. Parameters: $b = 0.15$, $\rho = 3$, $\varphi = 0.15$ unless varied, $\eta = 0$ in
    (a)--(d); $k_{\mathrm B}T = 0.01\,\omega_0$ in (a,b). All panels use the ground doublet with
    the microscopic arc couplings.
  }
  \label{fig:current-polarization}
\end{figure}

At $\alpha = 0$ the current polarization lies along the N\'eel axis,
$\bm P = P^z\hat{\bm z}$, with $P^z = (I_+ - I_-) / (I_+ + I_-)$ and $I_s$ the current carried
by N\'eel spin $s$. If the bias window contains only one level of a resolved doublet, that spin
carries the current and $P^z \to s$ [Fig.~\ref{fig:current-polarization}(a,b)], up to the tail of
the other resonance, which is negligible for
$|\Delta| \gg \bar\Gamma, k_{\mathrm B}T$. Once both resonances lie well inside the
bias window, their integrated weights [Eq.~\eqref{eq:sm-step-current}] give the
plateau currents. The polarization of the current plateau therefore equals the peak-height
asymmetry [Fig.~\ref{fig:current-polarization}(b)], and for a same-axis pair, for any ratio
$\Gamma_{0R}/\Gamma_{0L}$,
\begin{equation}
  \label{eq:sm-step-polarization}
  P^z = \mathcal A
      = \Pi^z
      = \frac{\mathcal I_+(\theta_j,\varphi) - \mathcal I_-(\theta_j,\varphi)}
             {\mathcal I_+(\theta_j,\varphi) + \mathcal I_-(\theta_j,\varphi)}.
\end{equation}
The last form holds for circular confinement, where $\Lambda_s$ and $\ell_{x,s}\ell_{y,s}$ in
Eq.~\eqref{eq:ground-state-hybridization-exact} are even in $s$ and only the angular integral of
Eq.~\eqref{eq:angular-integral-ground-state} distinguishes the two spins. To leading order in $b$
and for a narrow contact it reduces to $\tanh(b\rho^2\cos2\theta_j)$; a finite opening
averages the form factor over the arc and lowers $|P^z|$ [Fig.~\ref{fig:current-polarization}(c,d)].
The polarization vanishes at the nodes $\theta_j = \pm\pi/4$ and is odd in $b$, so it reverses with
the N\'eel vector. Lattice models of altermagnetic dots coupled to nonmagnetic leads likewise show
two-terminal spin filtering~\cite{Kirczenow_2026_Symmetry}.

Rashba coupling tilts $\bm P$ from the N\'eel axis at $O(\eta)$ and changes $|\bm P|$ only at
$O(\eta^2)$ [Fig.~\ref{fig:current-polarization}(e,f)]. This tilt differs from $\chi_j$ of
Sec.~\ref{sec:soc-doublet-block}: $\bm\Pi_j$ refers to the pseudospin of the dot doublet, $\bm P$
to the spin of the electron in the drain.

\section{Interacting transport formalism}
\label{sec:interacting-transport-formalism}

\subsection{Sequential dissipator and current}
\label{sec:sequential-dissipator-current}

We treat charging in the sequential-tunneling regime
$\bar\Gamma \ll k_{\mathrm B}T \ll U$ by retaining the complete reduced density matrix within
each charge sector. As in the main text, the projected gate energy $\varepsilon_g$ includes the
spin-averaged orbital energy of the lowest doublet. The lowest-doublet Anderson model has many-body
states $\ket{0}$, $\ket{s} = d_s^\dagger\ket{0}$ with $s = +, -$, and
$\ket{2} = d_+^\dagger d_-^\dagger\ket{0}$. For the constant-interaction Hamiltonian
$H_{\mathrm D} = \sum_s\varepsilon_s n_s + U n_+n_-$, an exactly degenerate doublet has
$E_0 = 0$, $E_s = \varepsilon_g$, and $E_2 = 2\varepsilon_g + U$.

For a basis-independent statement of the kinetic equation, diagonalize each positive
hybridization matrix as
\begin{equation}
  \label{eq:sm-gamma-eigenchannels}
  \Gamma_j = \sum_\kappa\gamma_{j\kappa} \ket{u_{j\kappa}}\!\bra{u_{j\kappa}}, \qquad d_{j\kappa}^\dagger
           = \sum_s \braket{s|u_{j\kappa}}d_s^\dagger.
\end{equation}
Writing $f_j(E) = \{1 + \exp[(E-\mu_j)/k_{\mathrm B}T_j]\}^{-1}$, the
electron-addition and electron-removal jump operators are
\begin{equation}
  \label{eq:sm-sequential-jumps}
  \begin{split}
    L_{j\kappa}^{+} & = \sqrt{\gamma_{j\kappa}}\sum_{a,b:\,N_b=N_a+1} \sqrt{f_j(E_b - E_a)}\, \ket{b}\bra{b}d_{j\kappa}^\dagger\ket{a}\bra{a}, \\
    L_{j\kappa}^{-} & = \sqrt{\gamma_{j\kappa}}\sum_{a,b:\,N_b=N_a+1} \sqrt{1 - f_j(E_b - E_a)}\, \ket{a}\bra{a}d_{j\kappa}\ket{b}\bra{b}.
  \end{split}
\end{equation}
This is the universal Lindblad equation~\cite{Nathan_2020_Universal, Nathan_2021_Erratum}, in
the form developed for quantum-dot
transport~\cite{Kirsanskas_2018_Phenomenological}, and it is the construction we
adopt at first order. We evaluate the dissipator with QmeQ~\cite{Kirsanskas_2017_QmeQ}
and include the principal-value Hamiltonian of
Sec.~\ref{sec:lamb-shift-ground-doublet}; time evolution is obtained by integrating
the resulting generator. The same generator also covers a finite level splitting,
such as one induced by an applied field.

The sequential part of the universal Lindblad equation is then
\begin{align}
  \label{eq:sm-sequential-dissipator}
  \left. \dot\varrho \right|_{\mathrm{diss}} & = \mathcal D[\varrho] = \sum_{j\kappa}\sum_{q=+,-} \mathcal D_{L_{j\kappa}^{q}}[\varrho], \\
  \mathcal D_L[\varrho]                      & = L\varrho L^\dagger - \frac12\{L^\dagger L,\varrho\}.
\end{align}
It contains real sequential tunneling only. The electrical current from lead $j$
into the dot is
\begin{equation}
  \label{eq:sm-sequential-current}
  I_j = e\sum_{\kappa}\left\{ \tr[L_{j\kappa}^{+}\varrho L_{j\kappa}^{+\dagger}] - \tr[L_{j\kappa}^{-}\varrho L_{j\kappa}^{-\dagger}] \right\},
\end{equation}
and the stationary solution obeys $I_L = -I_R$. For the spin-diagonal contacts used in main-text
Fig.~4, Eq.~\eqref{eq:sm-sequential-dissipator} reduces to the classical rate equation for the four
state populations.

Projected on the pseudospin, Eq.~\eqref{eq:sm-sequential-dissipator} shows why sequential tunneling
drops out during the dwell. With $\Gamma_j = \bar\Gamma_j(\bm1 + \bm\Pi_j \cdot \bm\sigma)$
as in Eq.~\eqref{eq:sm-effective-polarization}, let $P_N$ be the probability of charge sector $N$,
and abbreviate $f_j^0 = f_j(\varepsilon_g)$ and $f_j^2 = f_j(\varepsilon_g + U)$. For
$\varrho_1 = (P_1\bm1 + \bm S \cdot \bm\sigma) / 2$, projection onto the charge-one block gives
\begin{equation}
  \label{eq:sm-pseudospin-dissipator}
  \begin{split}
    \left. \dot{\bm S} \right|_{\mathrm{diss}} = \sum_j\bar\Gamma_j\Big\{ -[1 - f_j^0 + f_j^2]\bm S + \big[ 2f_j^0P_0 - 2(1 - f_j^2)P_2 + (f_j^0 + f_j^2 - 1)P_1 \big]\bm\Pi_j \Big\}.
  \end{split}
\end{equation}
The first part is the sequential-tunneling damping, while the last terms describe polarized loading
and escape.

\subsection{Lead-induced Lamb shift of the ground doublet}
\label{sec:lamb-shift-ground-doublet}

Virtual tunneling through the leads renormalizes the doublet. For the degenerate doublet,
$H_{\mathrm D} = \varepsilon_g(n_+ + n_-) + Un_+n_-$, and lead hybridization matrices
$(\Gamma_j)_{ss'} = 2\pi\sum_\nu t^\ast_{j\nu s}t_{j\nu s'}$ over orthogonal lead channels $\nu$,
the reduced density matrix obeys, to second order in the tunneling,
\begin{equation}
  \label{eq:lamb-shift-master-equation}
  \dot{\varrho} = -i\left[ H_{\mathrm D} + H_{\mathrm{LS}},\varrho \right] + \mathcal{D}[\varrho],
\end{equation}
with $\mathcal D$ the dissipator of Sec.~\ref{sec:sequential-dissipator-current} and
$H_{\mathrm{LS}}$ its principal-value counterpart, taken with the same universal-Lindblad
weights~\cite{Nathan_2020_Universal, Nathan_2021_Erratum}. Within the exactly degenerate doublet,
the transition energies coincide, so the geometric-mean correction vanishes.

In the wide-band limit, define
\begin{equation}
  \label{eq:lamb-shift-digamma-factor}
  \mathcal{W}_j(E) = \operatorname{Re}\psi\left( \frac{1}{2} + i\frac{E - \mu_j}{2\pi k_{\mathrm B}T_j} \right),
\end{equation}
where $\psi$ is the digamma function, as the wide-band, degenerate-doublet value of the
principal-value weight. A bandwidth-dependent constant has been omitted: within a
fixed charge sector it is proportional to the identity and therefore drops out of
Eq.~\eqref{eq:lamb-shift-master-equation}. In the singly occupied sector, virtual
tunneling through the empty state contributes
$\mathcal{W}_j(\varepsilon_g)\Gamma_j / (2\pi)$. The complementary path through the
doubly occupied state contributes
$\mathcal{W}_j(\varepsilon_g + U) [\tr(\Gamma_j)\bm{1} - \Gamma_j] / (2\pi)$. Their sum gives
\begin{equation}
  \label{eq:lamb-shift-charge-one-block}
  H_{\mathrm{LS}}^{(1)} = \sum_j\frac{1}{2\pi} \left\{ \mathcal{W}_j(\varepsilon_g)\Gamma_j + \mathcal{W}_j(\varepsilon_g + U) \left[ \tr(\Gamma_j)\bm{1} - \Gamma_j \right] \right\}.
\end{equation}

Decompose $H_{\mathrm{LS}}^{(1)} = \delta\varepsilon\bm1 + \bm B_{\mathrm{LS}} \cdot \bm\sigma/2$.
With
\begin{equation}
  \label{eq:sm-lead-factor}
  F_j
  = \mathcal W_j(\varepsilon_g) - \mathcal W_j(\varepsilon_g + U)
  = \mathcal P\!\!\int\! d\omega\, f_j(\omega) \left[ \frac{1}{\omega - \varepsilon_g} - \frac{1}{\omega - \varepsilon_g - U} \right],
\end{equation}
in which the omitted bandwidth constant cancels and the integral converges,
\begin{align}
  \delta\varepsilon                                & = \sum_j\frac{\bar\Gamma_j}{2\pi} [\mathcal W_j(\varepsilon_g) + \mathcal W_j(\varepsilon_g + U)], \\
  \label{eq:lamb-shift-vector} \bm B_{\mathrm{LS}} & = \sum_j\frac{F_j}{\pi}\bar\Gamma_j\bm\Pi_j.
\end{align}
The doublet splitting is $|\bm B_{\mathrm{LS}}|$. Each lead's contribution vanishes for
$\Gamma_j \propto \bm1$ or $F_j = 0$. Since $\operatorname{Re}\psi(1/2 + iy)$ is
even and strictly increasing in $|y|$, the latter occurs at $U = 0$ or
$\varepsilon_g = \mu_j - U/2$, at any temperature within this first-order wide-band
expression.

For spin-conserving contacts, only $B_z$ survives. This diagonal field commutes with the stationary
state and leaves the first-order stationary current unchanged; a torque requires hybridizations that
are not simultaneously diagonalizable. Rashba coupling, together with altermagnetism, supply them:
the dressed contacts of Sec.~\ref{sec:soc-doublet-block} carry the off-diagonal element $\zeta_j$,
which tilts each $\bm\Pi_j$ away from the N\'eel axis [Eq.~\eqref{eq:sm-effective-polarization}].
The torque on the source-prepared pseudospin, $\bm S(0) = \bm\Pi_L$ of
Eq.~\eqref{eq:pump-prepared-state}, is set by the component of $\bm B_{\mathrm{LS}}$ transverse to
$\bm\Pi_L$, which is invariant under the residual phase of the doublet basis. With Rashba coupling,
as in quantum-dot spin valves with ferromagnetic contacts, the field is generically
transverse to the polarization it acts on, so the doublet coherence must be kept, as
the universal Lindblad equation of Sec.~\ref{sec:sequential-dissipator-current}
does.

\subsection{Second-order real-time kernel}
\label{sec:second-order-kernel}

Cotunneling, which carries the conductance in the Coulomb valley (Sec.~\ref{sec:valley-zero-bias})
and damps the doublet precession (Sec.~\ref{sec:second-order-damping}), requires to work beyond the
first-order kernel. We compute the stationary state, the current, and the transient poles from the
Nakajima--Zwanzig kernel expanded to second order in the tunnel coupling,
$K = K^{(1)} + K^{(2)}$, with the real-time diagrammatic
technique~\cite{Schoeller_1994_Mesoscopic, Leijnse_2008_Kinetic,
Timm_2008_Tunneling, Koller_2010_Density} in the wide-band limit. Each contact enters through its
eigenchannels, Eq.~\eqref{eq:sm-gamma-eigenchannels}, so non-collinear polarizations appear only
through the complex overlaps $\braket{s|u_{j\kappa}}$. Because the doublet coherence is kept,
$K^{(2)}$ acts on populations and coherences and restricting it to the populations would drop terms
of the same order~\cite{Leijnse_2008_Kinetic, Koller_2010_Density}. For transients,
linearizing the memory, $K(\lambda) \simeq K(0) + \lambda K'(0)$, gives the generalized eigenproblem
\begin{equation}
  \label{eq:sm-linearized-poles}
  [\mathcal L_0 + K(0)]v = \lambda[\bm1 - K'(0)]v, \qquad \mathcal L_0
                         = -i[H_{\mathrm D},\, \cdot \,],
\end{equation}
where $\lambda K'^{(1)}(0)$ is kept because it is of the same order as $K^{(2)}$ for the slow poles,
$\lambda = O(\bar\Gamma)$.

\section{Pseudospin precession and charge readout}

\subsection{Pump--precess--probe response}
\label{sec:pump-precess-probe}

The protocol adapts the gate-pulsing scheme of Ref.~\cite{Hell_2015_Spin}, proposed for
ferromagnetic quantum-dot spin valves at fixed bias, to an unbiased dwell without ferromagnets.
A cycle consists of a load, a dwell, and a probe,
$X \in \{\mathrm{ld}, \mathrm{dw}, \mathrm{pr}\}$, of durations $t_{\mathrm{ld}}$,
$\tau$, and $t_{\mathrm{pr}}$. During the load part of the cycle, the level is placed inside the
bias window, $\mu_R < \varepsilon_g < \mu_L$, where only the source can fill it. The plunger
gate next moves the occupied dot into the singly occupied valley and the bias is
removed; real tunneling is suppressed there
[Eq.~\eqref{eq:sm-pseudospin-dissipator}], while the virtual processes in $H_{\mathrm{LS}}$ remain.
After the dwell part, the probe lifts the level above both chemical potentials, and the dot empties
into either lead; only exits through the drain transfer net charge. Repeating the sequence at rate
$f_{\mathrm{rep}}$ gives the dc current $I(\tau) = f_{\mathrm{rep}}Q_R(\tau)$, with $Q_R$ the charge
delivered to the drain per cycle, the pulse-induced-current readout of charge
qubits~\cite{Nakamura_1999_Coherent, Hayashi_2003_Coherent}.

\subsubsection{Ideal load and probe}
\label{sec:ideal-cycle}

In the ideal cycle, the source alone fills the empty dot, the probe is long, and the field due to
Lamb shift acts only during the dwell time. The electron enters the charge-one state
\begin{equation}
  \label{eq:pump-prepared-state}
  \varrho_1(0) = \frac{\Gamma_L}{\tr\Gamma_L},
\end{equation}
with $\bm S(0) = \bm\Pi_L$. During the probe it decays as
$\dot\varrho_1 = -\{\Gamma_L + \Gamma_R, \varrho_1\}/2$ and leaves through the drain with
probability $\tr(M_R\varrho_1)/\tr\varrho_1$, where
\begin{equation}
  \label{eq:pump-probe-readout}
  M_R = \int_0^\infty \dd t\, e^{-(\Gamma_L+\Gamma_R)t/2}\,\Gamma_R\,e^{-(\Gamma_L+\Gamma_R)t/2}, \qquad \{\Gamma_L + \Gamma_R, M_R\}
      = 2\Gamma_R.
\end{equation}
With $g = \sum_j\bar\Gamma_j$ and $\bm p = \sum_j\bar\Gamma_j\bm\Pi_j$, the anticommutator gives
$M_R = m_0 + \bm m \cdot \bm\sigma$ with $\bm m_\perp = \bar\Gamma_R\bm\Pi_{R \perp}/g$, where
$\perp$ denotes the part transverse to $\bm p$. Since $\bm p_\perp = 0$,
$\bar\Gamma_L\bm\Pi_{L \perp} = -\bar\Gamma_R\bm\Pi_{R \perp}$: the loaded pseudospin starts
antiparallel to $\bm m_\perp$.

At zero bias, $F_L = F_R \equiv F$ in Eq.~\eqref{eq:lamb-shift-vector}, so
$\bm B_{\mathrm{LS}} = F\bm p/\pi$. Equation~\eqref{eq:sm-pseudospin-dissipator} damps $\bm S$
isotropically at the rate $\gamma = \sum_j\bar\Gamma_j(1 - f_j^0 + f_j^2)$, loads it along
$\bm p$, and couples it to the charges only through $\bm p \cdot \bm S$. The transverse part
$\bm S_\perp$ thus precesses on its own at the frequency $\omega = |\bm B_{\mathrm{LS}}|$.
Without the field, real tunneling during the dwell only moves charge between the dwell and the
probe: each first-order process, entry or exit at $\varepsilon_g$ or $\varepsilon_g + U$, carries a
Fermi factor common to both leads, and by the anticommutator in Eq.~\eqref{eq:pump-probe-readout}
or its trace, the charge it delivers to the drain compensates the change it causes in the
drain-exit probability. The cycle charge therefore changes only through the torque of the field,
$\partial_\tau Q_R = e\,\bm m \cdot (\bm B_{\mathrm{LS}} \times \bm S)$. Integrating from its
field-free value at $\tau = 0$ gives
\begin{equation}
  \begin{split}
    Q_R(\tau) & = \frac{e\tr(M_R\Gamma_L)}{\tr\Gamma_L} + \frac{Q_\perp\omega}{\omega^2 + \gamma^2}\left[ \omega - e^{-\gamma\tau}\left( \omega\cos\omega\tau + \gamma\sin\omega\tau \right) \right], \\
    Q_\perp   & = \frac{e\bar\Gamma_L\bar\Gamma_R^2\left| \bm\Pi_L \times \bm\Pi_R \right|^2}{g|\bm p|^2}.
  \end{split}
\end{equation}
In the form of main-text Eq.~(12), $Q_1 = Q_\perp/\sqrt{1 + \gamma^2/\omega^2}$ and
$\Phi = \pi - \arctan(\gamma/\omega)$. The bracket is nonnegative, so the field only
raises $Q_R$ above its field-free value. The amplitude is set by the contact
polarizations alone and vanishes when they are collinear, while the field sets only
$\omega$.

\subsubsection{Nonideal load and probe}

For a load of finite duration and selectivity, and a field acting in every stage, we retain all
charge sectors and evolve each stage with $\mathcal L_X$ of
Eq.~\eqref{eq:lamb-shift-master-equation}, evaluated at the degenerate level $\varepsilon_X$, bias
$\mu_{L,R} = \pm V_X/2$, and common temperature $T$. We define the drain particle-current
functional $\mathcal J_X = -I_R/e$ from Eq.~\eqref{eq:sm-sequential-current} and the integrated
propagator $\mathcal K_X(t) = \int_0^t\dd t'\,e^{\mathcal L_X t'}$. The probe remains long: it
resets the dot to its stationary state, empty up to thermal occupation, so successive cycles are
independent. Since a zero-bias stationary state carries no current, the drain-exit functional
$\mathcal M_R = \lim_{t \to \infty}\mathcal J_{\mathrm{pr}}\mathcal K_{\mathrm{pr}}(t)$ converges,
and the probe delivers the charge $e\,\mathcal M_R\varrho$ from the state $\varrho$ it starts
with. On a singly occupied state, $\mathcal M_R$ gives the drain-exit probability and reduces
to $\tr(M_R\varrho_1)$ when the field during the probe is neglected. Starting from
$\varrho_\emptyset = \ket{0}\!\bra{0}$, the load prepares
$\varrho_{\mathrm{ld}} = e^{\mathcal L_{\mathrm{ld}}t_{\mathrm{ld}}}\varrho_\emptyset$, and the
cycle charge is
\begin{equation}
  \label{eq:sm-cycle-charge}
  \frac{Q_R(\tau)}{e}
  = \mathcal J_{\mathrm{ld}}\mathcal K_{\mathrm{ld}}(t_{\mathrm{ld}})\varrho_\emptyset
    + \mathcal J_{\mathrm{dw}}\mathcal K_{\mathrm{dw}}(\tau)\varrho_{\mathrm{ld}}
    + \mathcal M_R\,e^{\mathcal L_{\mathrm{dw}}\tau}\varrho_{\mathrm{ld}},
\end{equation}
the drain charge of the load, the dwell, and the probe. Write
$e^{\mathcal L_{\mathrm{dw}}\tau} = \sum_k e^{\lambda_k\tau}\mathcal P_k$, with eigenvalues
$\lambda_k$, spectral projectors $\mathcal P_k$, and $\lambda_0 = 0$ for the stationary state,
which again carries no current, $\mathcal J_{\mathrm{dw}}\mathcal P_0 = 0$, so
\begin{equation}
  \label{eq:sm-cycle-modes}
  \frac{Q_R(\tau)}{e}
  = \frac{Q_R(\infty)}{e} + \sum_{k \ne 0}c_k\,e^{\lambda_k\tau}, \qquad c_k
  = \left[ \mathcal M_R + \frac{\mathcal J_{\mathrm{dw}}}{\lambda_k} \right]\mathcal P_k\varrho_{\mathrm{ld}}.
\end{equation}
This is main-text Eq.~(12): the complex pair $\lambda_\pm = -\gamma \pm i\omega$ gives
$Q_1e^{i\Phi} = 2ec_+$, and $Q_{\mathrm{bg}} = Q_R(\infty)$, because the real modes carry no
weight. As in Sec.~\ref{sec:ideal-cycle}, $\bm S_\perp$ precesses on its own, giving
$\lambda_\pm$, while the real modes involve only the charges and $\bm S \cdot \bm p$, on which the
field exerts no torque. For a field-free probe, the compensation of Sec.~\ref{sec:ideal-cycle}
reads
$\mathcal J_{\mathrm{dw}} = -\mathcal M_R(\mathcal L_{\mathrm{dw}} - \mathcal L_{\mathrm{LS}})$,
with $\mathcal L_{\mathrm{LS}} = -i[H_{\mathrm{LS}},\, \cdot \,]$, so
$c_k = \mathcal M_R\,\mathcal L_{\mathrm{LS}}\mathcal P_k\varrho_{\mathrm{ld}}/\lambda_k$ and the
real modes drop out, also with a field during the probe, which is again parallel to $\bm p$. Only a
probe too short to empty the dot leaves small real-mode weights. The coefficients depend on the
load and the probe through $\varrho_{\mathrm{ld}}$ and $\mathcal M_R$, while the
poles belong to the dwell alone.

Main-text Fig.~5(d) uses the contacts and lead parameters of its caption
($b = 0.15$, $\rho = 3$, $\varphi = 0.15$, $\eta = 0.15$, $\theta_L = 0$, $\theta_R = \pi$,
$\bar\Gamma = 0.01U$, $k_{\mathrm B}T = 0.05U$) and the stages
$(\varepsilon_{\mathrm{ld}}, V_{\mathrm{ld}}) = (0, 0.4U)$, with the level at the center of the bias
window, for $\bar\Gamma t_{\mathrm{ld}} = 1$;
$(\varepsilon_{\mathrm{dw}}, V_{\mathrm{dw}}) = (-U/4, 0)$; and
$(\varepsilon_{\mathrm{pr}}, V_{\mathrm{pr}}) = (U/2, 0)$ with a long probe. The load is not long
compared with $1/\bar\Gamma$ and is not ideally selective, so
$\varrho_{\mathrm{ld}}$ retains an empty part and a pseudospin between
Eq.~\eqref{eq:pump-prepared-state} and $\varrho_{\mathrm{ld}}^{\mathrm{ss}}$.

\subsection{Second-order damping of the precession}
\label{sec:second-order-damping}

\begin{figure}[t]
  \centering
  \includegraphics[width=\linewidth]{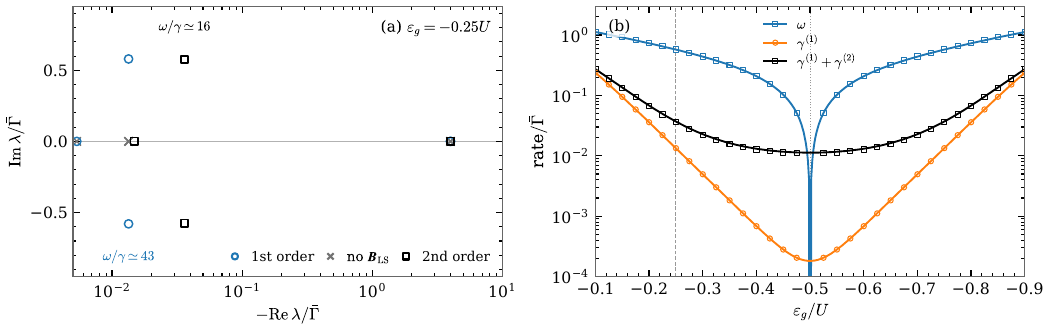}
  \caption{
    Precession poles of the singly occupied doublet, same-axis contacts at the working point of
    main-text Fig.~5, zero bias. (a) Nonstationary poles $\lambda$ at the dwell point
    $\varepsilon_g = -U/4$: first order with (circles) and without (crosses)
    $\bm B_{\mathrm{LS}}$, and second order including the memory of the kernel (squares). (b) The
    precession pair across the valley: frequency $\omega$ and decay rate at first order,
    $\gamma^{(1)}$, and at second order, $\gamma^{(1)} + \gamma^{(2)}$. Markers: poles of the
    first- and second-order kernels; lines: Eqs.~\eqref{eq:sm-equilibrium-poles}--%
    \eqref{eq:sm-complete-second-order-field}. Dotted: $\varepsilon_g = -U/2$, where the field
    vanishes; dashed: the dwell point of (a).
  }
  \label{fig:precession-poles}
\end{figure}

Cotunneling, the leading real process in the valley, is not suppressed by the dwell.
With the second-order kernel of Sec.~\ref{sec:second-order-kernel} and the poles of
Eq.~\eqref{eq:sm-linearized-poles}, the precession frequency shifts only by an amount exponentially
small in the valley while the decay rate grows [Fig.~\ref{fig:precession-poles}(a)]: the readout
stays underdamped but resolves fewer oscillations. Finite pulse shapes and environmental noise, such
as gate-voltage fluctuations, are omitted here; they would reduce the visibility
further and modify the envelope of the decaying oscillations in main-text Fig.~5.

The growth of the decay rate follows from the golden rule. Write $p = |\bm p|$. At zero bias, with
$a_0 = 1/(\mu - \varepsilon_g)$ and $a_2 = 1/(\varepsilon_g + U - \mu)$, both small compared with
$1/k_{\mathrm B}T$ inside the valley, $\mu - U < \varepsilon_g < \mu$, an electron transferred from
channel $k\kappa'$ to channel $j\kappa$ of Eq.~\eqref{eq:sm-gamma-eigenchannels}, through the empty
or the doubly occupied virtual state, acts on the singly occupied doublet with the jump operator
\begin{equation}
  \label{eq:sm-cotunneling-jumps}
  L_{j\kappa,k\kappa'} = \sqrt{\frac{\gamma_{j\kappa}\gamma_{k\kappa'}k_{\mathrm B}T}{2\pi}} \left[ (a_0 + a_2)\ket{u_{k\kappa'}}\!\bra{u_{j\kappa}} - a_2\braket{u_{j\kappa}|u_{k\kappa'}} \right].
\end{equation}
Here $k_{\mathrm B}T = \int\dd\omega\,f(\omega)[1 - f(\omega)]$ is the phase space of an
elastic transfer at zero bias, and the energy dependence of the denominators is
neglected. The identity part of $L_{j\kappa,k\kappa'}$ gives no net relaxation after
summing the equilibrium channel pairs. Summing $\mathcal D_L$ over all channel pairs with
$\sum_\kappa\gamma_{j\kappa}\ket{u_{j\kappa}}\!\bra{u_{j\kappa}} = \Gamma_j$ gives
$\dot{\bm S}|_{\mathrm{cot}} = -R\bm S$ with no source term and
\begin{equation}
  \label{eq:sm-cotunneling-relaxation}
  R = \frac{k_{\mathrm B}T}{\pi}(a_0 + a_2)^2\left[ g^2\bm 1 - \bm p\bm p^{\mathsf T} \right].
\end{equation}
$R$ is axially symmetric about $\bm p$, so, as in Sec.~\ref{sec:ideal-cycle}, $\bm S_\perp$
precesses on its own, and it decays at
$\gamma_{\mathrm{cot}} = (k_{\mathrm B}T/\pi)(a_0 + a_2)^2g^2$ for any contact polarizations.
Together with the first-order rate $\gamma^{(1)}$, the $\gamma$ of Sec.~\ref{sec:ideal-cycle},
the complex pair within this cotunneling approximation is
$-(\gamma^{(1)} + \gamma_{\mathrm{cot}}) \pm i|\bm B_{\mathrm{LS}}|$. In the
weak-coupling deep-valley regime
$g \ll k_{\mathrm B}T \ll \min(\mu - \varepsilon_g,\varepsilon_g + U - \mu)$,
$\gamma_{\mathrm{cot}} \propto \bar\Gamma^2k_{\mathrm B}T/U^2$ at fixed $\varepsilon_g/U$.
Where the sequential rate is negligible, the quality factor saturates at
\begin{equation}
  \label{eq:sm-quality-factor-cotunneling}
  \frac{\omega}{\gamma} \simeq  \frac{|\bm B_{\mathrm{LS}}|}{\gamma_{\mathrm{cot}}}
                        =       \frac{|F|\,p}{g^2k_{\mathrm B}T(a_0 + a_2)^2}
                        \propto \frac{U^2}{\bar\Gamma k_{\mathrm B}T}.
\end{equation}
Away from the field node, this permits underdamped precession at weak coupling.

The full second-order poles also include virtual corrections and the first-order kernel's
Laplace derivative. Away from the deep valley the energy dependence of the denominators
matters; it is resummed by the finite-part cotunneling integral. Set $E_0 = \varepsilon_g - \mu$,
$E_2 = E_0 + U$, and $\mathcal W_n = \mathcal W_j(\mu + E_n)$ for $n = 0,2$, with $\mathcal W_j$ of
Eq.~\eqref{eq:lamb-shift-digamma-factor}, the same for both leads at common $\mu$ and $T$. Primes
denote energy derivatives. Then
\begin{equation}
  \label{eq:sm-cotunneling-finite-part}
  \begin{split}
    \mathcal I & = \operatorname{Fp}\!\int\dd x\,f(x)[1 - f(x)] \left[ \frac{1}{x - E_0} - \frac{1}{x - E_2} \right]^2               \\
               & = -k_{\mathrm B}T\left[ \mathcal W_0'' + \mathcal W_2'' - 2\frac{\mathcal W_0' - \mathcal W_2'}{E_0 - E_2} \right],
  \end{split}
\end{equation}
where $f(x) = (e^{x/k_{\mathrm B}T} + 1)^{-1}$ and $\operatorname{Fp}$ denotes the Hadamard finite
part, removing the on-shell sequential divergence, which replaces $k_{\mathrm B}T(a_0 + a_2)^2$ in
$\gamma_{\mathrm{cot}}$. This regularized cotunneling estimate alone does not
include every finite-temperature second-order
contribution~\cite{Koller_2010_Density, Timm_2011_Time}. For the bare-degenerate
doublet at common lead temperature and chemical potential, with energy-independent hybridizations
in the wide-band limit, the remaining corrections also admit closed forms. Define $f_n = f(E_n)$,
so that $f_n' = -f_n(1 - f_n)/k_{\mathrm B}T$. In addition to $\mathcal I$ of
Eq.~\eqref{eq:sm-cotunneling-finite-part}, introduce
\begin{equation}
  \label{eq:sm-pole-thermal-factor}
  C = (1 - f_0 - f_2)\left[ \frac{\mathcal W_0' + \mathcal W_2'}{2} + \frac{F}{U} + \frac{(f_0 - f_2)F}{2k_{\mathrm B}T} \right].
\end{equation}
The transverse poles, expanded at fixed nonzero temperature through second order in the
hybridization (fourth order in tunneling amplitudes), are
\begin{align}
  \label{eq:sm-equilibrium-poles}
  \lambda_s                                                & = -\gamma^{(1)} - \gamma^{(2)} + s\,i[B^{(1)} + B^{(2)}] + O(\bar\Gamma^3),\qquad s = \pm1,  \\
  \gamma^{(1)}                                             & = g(1 - f_0 + f_2),\qquad B^{(1)} = pF/\pi,\nonumber                                         \\
  \label{eq:sm-complete-second-order-damping} \gamma^{(2)} & = \frac{g^2\mathcal I - (g^2 - p^2)C}{\pi},                                                  \\
  \label{eq:sm-complete-second-order-field} B^{(2)}        & = gp\left[ (1 - 2f_0)f_0' + (1 - 2f_2)f_2' - \frac{2k_{\mathrm B}T}{U}(f_0' - f_2') \right].
\end{align}
The field amplitudes are signed along $\bm p/p$; the precession frequency is
$|B^{(1)} + B^{(2)}|$. They follow from expanding the pole of each transverse coherence, a
scalar sector of the kernel by axial symmetry, through second order. At
particle--hole symmetry, $C = B^{(2)} = 0$. Deep in the valley these terms are thermally
suppressed, while $\gamma^{(2)}$ reduces to $\gamma_{\mathrm{cot}}$ and gives the quality factor of
Eq.~\eqref{eq:sm-quality-factor-cotunneling}.

\section{Stationary transport through the degenerate doublet}

\begin{figure}[t]
  \centering
  \includegraphics[width=\linewidth]{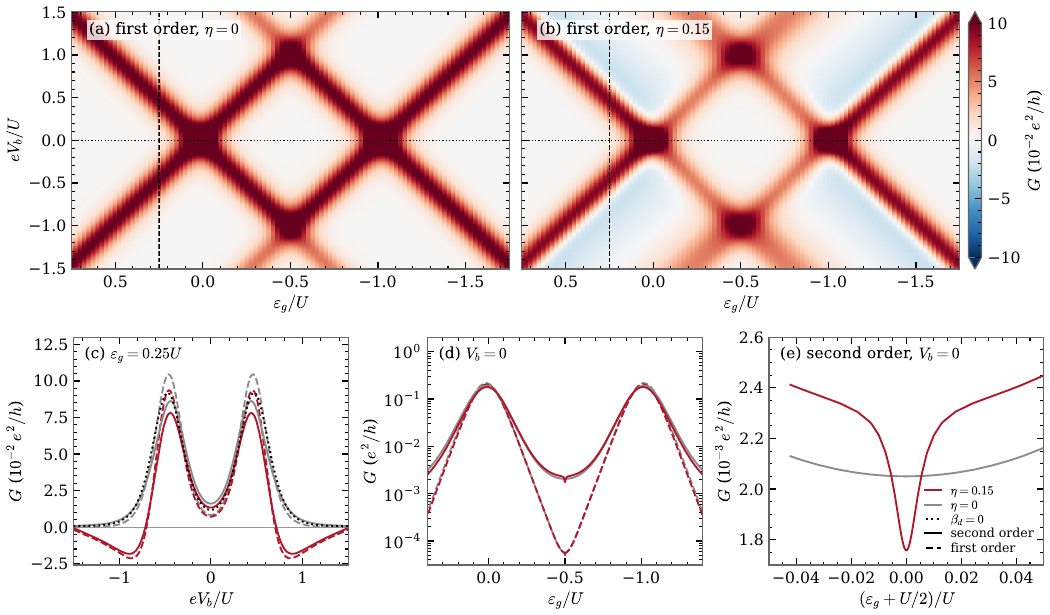}
  \caption{
    Transport through the degenerate doublet, same-axis contacts at the working point of main-text
    Fig.~5. (a),(b) First-order $dI/dV$ against gate level and bias for collinear ($\eta = 0$) and
    non-collinear ($\eta = 0.15$) polarizations; the gate axis runs as in main-text Fig.~4, and
    the dashed and dotted lines mark the cuts of (c) and (d). (c) $dI/dV$ against bias at
    $\varepsilon_g = 0.25\,U$ (identical at $-1.25\,U$ by particle--hole symmetry): solid, second
    order; dashed, first order; dotted, $\beta_d = 0$ at $\eta = 0.15$, second order. (d) Linear
    conductance across the valley. (e) Neighborhood of the particle--hole point, second order.
    Parameters: $b = 0.15$, $\rho = 3$, $\varphi = 0.15$, $\Delta = 0$,
    $k_{\mathrm B}T / U = 0.05$ and $\bar\Gamma / U = 0.01$. First order keeps the doublet
    coherence.
  }
  \label{fig:degenerate-transport}
\end{figure}

We evaluate the stationary current with the first- and second-order real-time kernels of
Sec.~\ref{sec:second-order-kernel}, retaining the doublet coherences at both orders
(Fig.~\ref{fig:degenerate-transport}).

\subsection{Negative differential conductance from non-collinear polarizations}
\label{sec:noncollinear-ndc}

Outside the blockade the non-collinear polarizations have a consequence already at first order.
Beyond the first conductance step the current decreases with bias, and $dI/dV$ turns negative in
wedges that flank the diamonds [Fig.~\ref{fig:degenerate-transport}(b)]; the cut in
Fig.~\ref{fig:degenerate-transport}(c) shows it at both orders. It is absent for collinear
polarizations, $\eta = 0$ [Fig.~\ref{fig:degenerate-transport}(a)], and for $\beta_d = 0$, where
time-reversal symmetry leaves each contact unpolarized [Fig.~\ref{fig:degenerate-transport}(c),
dotted]. It therefore requires both the altermagnetic polarization of the contacts and its
spin--orbit tilt. Unlike the negative differential conductance of main-text Fig.~4, which comes
from the unequal rates of a split doublet, it appears with the doublet degenerate.

\subsection{Zero-bias conductance in the Coulomb valley}
\label{sec:valley-zero-bias}

Inside the valley the first-order conductance is exponentially suppressed, while
cotunneling fills it [Fig.~\ref{fig:degenerate-transport}(d)]; the generic cotunneling
conductance, $\propto 1/\varepsilon_g^2 + 1/(\varepsilon_g + U)^2$, has its minimum at $-U/2$ but
varies on the scale of $U$.

For $\eta = 0$ the polarizations are collinear: the exchange field is parallel to both and to the
pseudospin loaded by the source, so it exerts no torque. Rashba coupling tilts the two
polarizations in opposite directions. The source then populates a pseudospin state weakly
coupled to the drain, suppressing the current. The component of $\bm B_{\mathrm{LS}}$ transverse to
that pseudospin rotates it toward the strongly coupled channel, raising the cotunneling
conductance above its $\eta = 0$ value [Fig.~\ref{fig:degenerate-transport}(e)], as for
ferromagnet-contacted dots in sequential tunneling~\cite{Konig_2003_Interaction, Braun_2004_Theory}
and in the Coulomb valley~\cite{Weymann_2007_Cotunneling}. At zero bias both leads carry the same
factor $F$, so the field has a fixed direction and a magnitude proportional to $F$, which vanishes
at $\varepsilon_g = -U/2$ for arbitrary contact matrices in the degenerate, single-orbital wide-band
model at equal lead temperatures. There the rotation stops and the conductance drops to its blocked
value, below the collinear one, giving a narrow dip pinned at the particle--hole point
[Fig.~\ref{fig:degenerate-transport}(e)]. It extends over the detuning at which
$|\bm B_{\mathrm{LS}}|$ falls below the cotunneling relaxation rate $\gamma_{\mathrm{cot}}$ of
Sec.~\ref{sec:second-order-damping},
$|\varepsilon_g + U/2| \lesssim \gamma_{\mathrm{cot}}/|\partial_{\varepsilon_g}\bm B_{\mathrm{LS}}| = 4k_{\mathrm B}T(\sum_j\bar\Gamma_j)^2/(U|\bm p|)$,
of order $\bar\Gamma k_{\mathrm B}T/U$, and widens with temperature and tunnel
coupling accordingly (not shown); collinear polarizations show none. N\'eel reversal
leaves it unchanged, as Onsager reciprocity requires of the linear conductance. A bare splitting
$\Delta$ adds to $B^z_{\mathrm{LS}}$, which lies along the N\'eel axis for the same-axis pair, and
moves the zero of the total field, and with it the dip, to the gate level where
$B^z_{\mathrm{LS}} = -\Delta$, if the valley contains one; energy-dependent hybridization need not
preserve the node.

\section{Orthogonal contacts: a magnet-free spin valve}
\label{sec:orthogonal-contacts}

\begin{figure}[t]
  \centering
  \includegraphics[width=\linewidth]{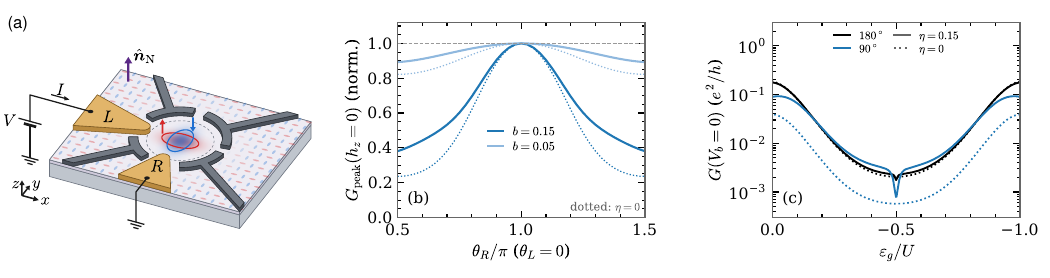}
  \caption{
    Orthogonal contacts and the spin-valve geometry. (a) Device with the drain on the $y$
    axis: the two contacts favor opposite N\'eel spins. (b) Zero-field peak conductance
    as the drain rotates from the same-axis position, $\theta_R = \pi$, to the orthogonal
    positions, normalized to its same-axis value; dotted curves: $\eta = 0$. (c)
    Zero-bias conductance across the Coulomb valley at second order for the same-axis
    ($180^\circ$) and orthogonal ($90^\circ$) pairs; solid, $\eta = 0.15$; dotted, $\eta = 0$.
    Panel (b): microscopic Landauer transport, $\rho = 3$, $\varphi = 0.15$,
    $k_{\mathrm B}T = 0.01\,\omega_0$, $\eta = 0.15$ and $b = 0.15$ unless indicated.
    Panel (c): working point of Fig.~\ref{fig:degenerate-transport}.
  }
  \label{fig:orthogonal-contacts}
\end{figure}

The main text uses a same-axis contact pair throughout. Here we consider the orthogonal case,
$\theta_L = 0$ and $\theta_R = \pi / 2$ [Fig.~\ref{fig:orthogonal-contacts}(a)]. Since the $d$-wave
form factor reverses under a $90^\circ$ rotation of the contact,
Eq.~\eqref{eq:sm-effective-polarization} gives
\begin{equation}
  \label{eq:sm-antiparallel-configuration}
  \Pi^z\!\left( \theta + \pi / 2 \right) = -\Pi^z\!\left( \theta \right) \qquad \left( \alpha = 0 \right),
\end{equation}
so two ordinary metallic reservoirs on orthogonal principal axes realize the projected analogue of
an antiparallel quantum-dot spin valve, with no ferromagnetic reservoir or applied field. ``Spin
valve'' refers to the effective coupling within the lowest doublet; the reservoirs remain
unpolarized. The relation survives a finite contact width
(Sec.~\ref{sec:derivation-of-the-hybridization-matrix}), while the selectivity vanishes at
$\beta_d = 0$. At the single-particle level the valve suppresses transport relative to a same-axis
pair by $1 - \Pi^2 = \operatorname{sech}^2 A$. The suppression grows exponentially with the
confinement energy set by the gates. A same-axis pair thus reports the selectivity as unequal peak
heights and a polarized current at full transmission, an orthogonal pair as a
suppressed, unpolarized current.

We now evaluate the resonant transmission of the lowest doublet directly from the Pauli
decomposition of Eq.~\eqref{eq:sm-effective-polarization}, without assuming that the two contacts
share an eigenbasis. Let the doublet be degenerate at $\varepsilon_g$ and let both contacts have
the same mean coupling, $\bar{\Gamma}_L = \bar{\Gamma}_R \equiv \bar{\Gamma}$, which
for geometrically identical $x / y$ contacts with equal bare tunnel strengths
follows from $C_4\mathcal{T}$ because $\bar{\Gamma}_j = \Gamma_{0j}\cosh A_j$ is even in
$A_j$. Evaluating Eq.~\eqref{eq:retarded-greens-function} at $E = \varepsilon_g$ gives
$G^R = -2iS^{-1}$ with $S = \Gamma_L + \Gamma_R$, so the Caroli
formula~\eqref{eq:caroli-formula} becomes
\begin{equation}
  \label{eq:sm-peak-transmission-matrix}
  T_{\mathrm{peak}} = 4\,\tr\!\left[ \Gamma_L S^{-1}\Gamma_R S^{-1} \right].
\end{equation}
For equal polarization magnitudes $\Pi_L = \Pi_R \equiv \Pi$ and relative angle
$\chi_{LR}$ of Eq.~\eqref{eq:sm-relative-angle}, evaluating the Pauli-matrix trace gives
\begin{equation}
  \label{eq:sm-valve-transmission}
  T_{\mathrm{peak}} = \frac{2\left( 1 - \Pi^2 \right)}{1 - \Pi^2\cos^2\left( \chi_{LR} / 2 \right)}.
\end{equation}
The two collinear limits are the familiar ones: $\chi_{LR} = 0$ gives the unitary value
$T_{\mathrm{peak}} = 2$ for any selectivity, because a parallel pair transmits both eigenchannels
symmetrically, while $\chi_{LR} = \pi$ gives $2(1 - \Pi^2)$, reproducing
Eq.~\eqref{eq:sm-thermal-valve-ratio}.

At $\eta = 0$ the transmission peaks of the two spin species are equal, with the same height and
weight. Rashba coupling leaves the equality of the two peaks intact to leading order. The
antiunitary $C_4\mathcal{T}$ symmetry maps a contact at $\theta$ onto one at $\theta + \pi / 2$ and
reverses the longitudinal pseudospin component. It therefore enforces $\Pi_{R,z} = -\Pi_{L,z}$ at
every Rashba strength, while the in-plane components acquired at $O(\eta)$ are rotated by
$\pi / 2$ between the two contacts. Peaks resolved along the N\'eel axis see only the $z$
components, so the projected model predicts equal heights for the two spins at any $\eta$, and the
conductance at $\pm h_z$ coincides exactly. The in-plane components do enter at zero field, where
the degenerate doublet transmits coherently through contacts with non-collinear
polarizations. There the valve opens: the peak conductance of the orthogonal pair rises above its
$\eta = 0$ value ($\propto \operatorname{sech}^2A$), as $\eta$ grows
[Fig.~\ref{fig:orthogonal-contacts}(b)], in line with
Eq.~\eqref{eq:sm-valve-transmission} for $\chi_{LR} < \pi$. The coherent zero-field
regime is confined to $|h_z| \lesssim \bar\Gamma$; for any resolvable field the two levels transmit
independently.

The valve also acts on cotunneling in the Coulomb valley [Fig.~\ref{fig:orthogonal-contacts}(c)].
For $\eta = 0$ the orthogonal pair conducts less than the same-axis pair across the valley, and
Rashba coupling raises its conductance above its $\eta = 0$ value at every gate level. The zero-bias
dip of Sec.~\ref{sec:valley-zero-bias} appears for both pairs, pinned at $\varepsilon_g = -U/2$, and
is deeper for the orthogonal one. At the dip, where the exchange field vanishes, the orthogonal
conductance stays above its $\eta = 0$ value, whereas the same-axis one falls below it.

\bibliography{amdot.bib}